\documentclass{elsarticle}

\usepackage{amsmath}
\usepackage{bm}
\usepackage{svg}
\usepackage{amssymb}
\usepackage{lipsum}
\usepackage{listings}
\usepackage{gensymb}
\usepackage{fancyhdr}
\usepackage{graphicx}
\usepackage{url}
\usepackage{lineno}
\usepackage{xcolor}
\usepackage{color}
\usepackage{graphicx}
\usepackage[breaklinks]{hyperref}

\usepackage{geometry}
\newcommand{\fracnot}[3]{\mathcal{F}_{#1}^{#2,#3}}

\begin{document}
\begin{frontmatter}

\title{Squeeze: Efficient Compact Fractals for Tensor Core GPUs}

\address[a]{Instituto de Informática, Facultad de Ciencias de la Ingenier\'ia, Universidad Austral de Chile.}

\address[b]{Computer Science Department (DCC), University of Chile.}

\cortext[cor2] {Corresponding author.}

\author[a]{Felipe A. Quezada}

\author[a]{Crist\'obal A. Navarro\corref{cor2}}
\ead{cnavarro@inf.uach.cl}

\author[b]{Nancy Hitschfeld}

\author[b]{Benjamin Bustos}


\begin{abstract}
This work presents Squeeze, an efficient compact fractal processing scheme for tensor core GPUs. By combining discrete-space transformations between compact and expanded forms, one can do data-parallel computation on a fractal with neighborhood access without needing to expand the fractal in memory. The space transformations are formulated as two GPU tensor-core accelerated thread maps, $\lambda(\omega)$ and $\nu(\omega)$, which act as compact-to-expanded and expanded-to-compact space functions, respectively. The cost of the maps is $\mathcal{O}(\log_2 \log_s(n))$ time, with $n$ being the side of a $n \times n$ embedding for the fractal in its expanded form, and $s$ the linear scaling factor. The proposed approach works for any fractal that belongs to the Non-overlapping-Bounding-Boxes (NBB) class of discrete fractals, and can be extended to three dimensions as well. Experimental results using a discrete Sierpinski Triangle as a case study shows up to $\sim12\times$ of speedup and a memory reduction factor of up to $\sim 315\times$ with respect to a GPU-based expanded-space bounding box approach. 
These results show that the proposed compact approach will allow the scientific community to efficiently tackle problems that up to now could not fit into GPU memory.

\end{abstract}


\end{frontmatter}

\section{Introduction}
\label{sec:intro}
Many natural phenomena exhibit fractal like features in their structure, such as vegetation growth \cite{Oppenheimer:1986:RTD:15886.15892,Palmer1988}, terrain formation 
\cite{MILNE198867,4767591}, molecular dynamic patterns \cite{rothemund2004}, blood vessels generation \cite{PhysRevLett.90.118101}, among many other examples.
Fractals structures exhibit self-similarity, a property where the whole structure's information is present at different levels of scale. Mathematical definitions of fractal geometry have been formulated in order to model and simulate natural phenomena that cannot be easily explained in terms of traditional Euclidean geometry. 

\begin{figure}[ht!]
\centering
\includegraphics[scale=0.30]{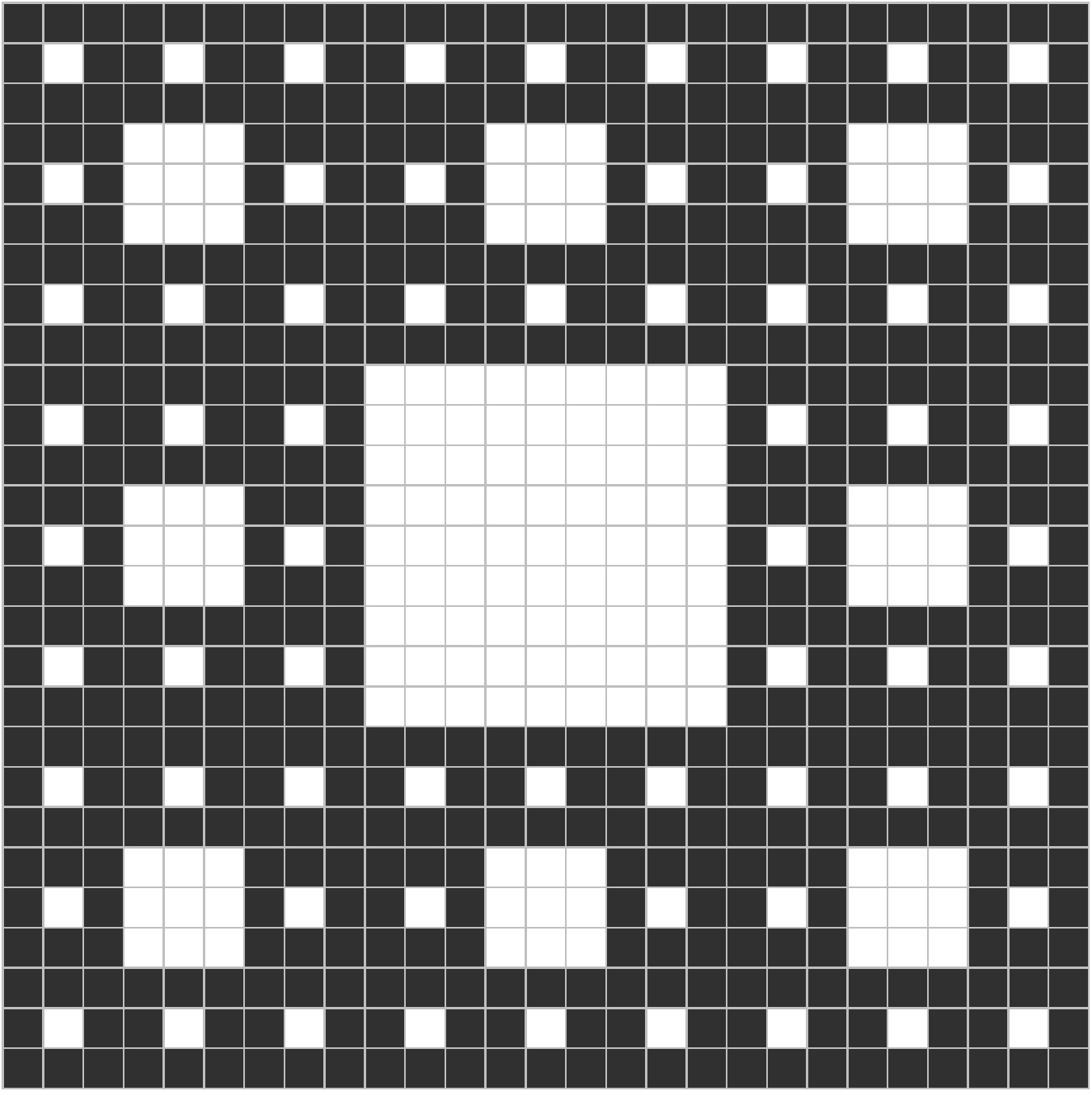}
\caption{The Sierpiński Carpet embedded in a discrete euclidean domain of $27 \times 27$ elements.}
\label{fig:carpet-fractal-embedded}
\end{figure}

One known approach often employed in parallel simulations is to use a discrete \textit{embedded} representation of the fractal, where it is contained inside a \textit{bounding-box} embedding in expanded form. 
Although this representation indeed simplifies the mapping of parallel threads onto data elements and the exploration of its neighbors, it sacrifices compute resources as it discards threads that fall outside the region of interest at run-time, as well as memory by having to store the whole embedding instead of just the fractal. 
Figure \ref{fig:carpet-fractal-embedded} depicts an example fractal embedded in a bounding-box. 

This parallel resource and memory problem is not specific to just one fractal, but to all discrete fractals handled this way. 
In particular, here we focus on the Non-overlapping Bounding-Boxes (NBB) class of fractals~\cite{NAVARRO2020158}, which satisfies two properties: i) the smallest level of the fractal occupies one unit of discrete space and from that point on, it can only scale up, and ii) each fractal has a unique transition function that takes the fractal in its current scale level, and replicates it in space to generate the fractal at the next scale level. In the NBB class, we assume that replicas can be translated, but cannot rotate neither overlap with each other.

\subsection{The two problems with embedded fractals}
In the embedded representation, as the fractal gets larger, 
the number of fractal data-elements will become asymptotically smaller than the number of non-fractal elements (the empty spaces or \textit{holes} of the embedding), bringing up two problems:
\begin{itemize}
    \item \textbf{[parallel efficiency]} P1: The number of computational resources (threads from the grid) mapped to the problem will grow in terms of the bounding-box (embedding space), and not in terms of the number of elements of the fractal which is what is actually needed. With small fractals the problem is not so dramatic, but as the fractal grows, this difference between fractal and empty elements gets larger.
    \item \textbf{[memory usage]} P2: The memory usage will increase in terms of the bounding-box, and not in terms of the fractal, putting a very early limit on the largest problem size that fits in the GPU.
\end{itemize}

\begin{figure}[ht!]
\centering
\includegraphics[scale=0.14]{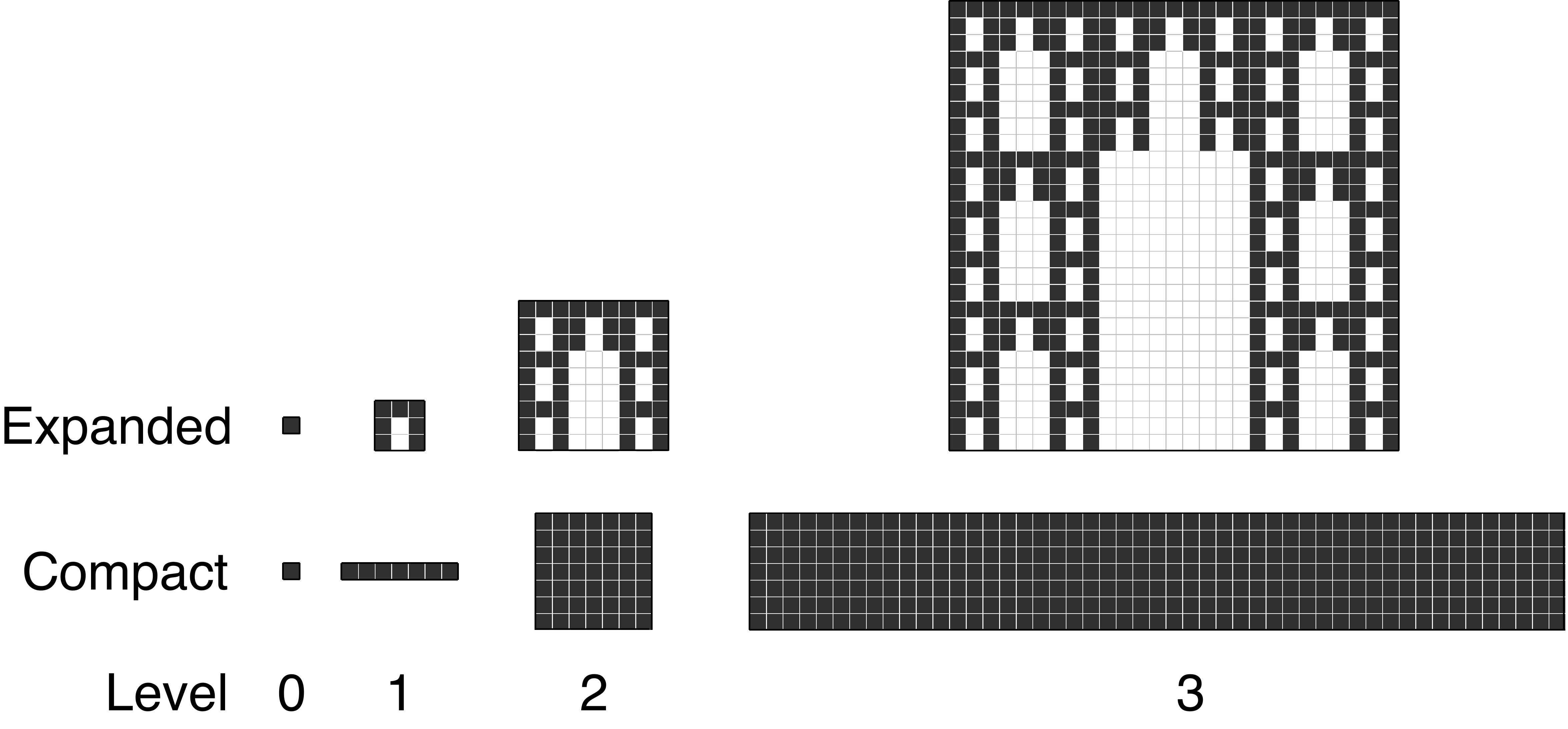}
\caption{The construction of the empty-bottles NBB fractal.}
\label{fig:nbb-construction}
\end{figure}

The research question is whether there exists an efficient GPU scheme for solving P1 and P2. 
To achieve this, the fractal must be processed in its compact form and not in the expanded embedded one. Figure \ref{fig:nbb-construction} shows an example NBB fractal, with its compact representation below, at each level.

\begin{figure}[ht!]
\centering
\includegraphics[scale=0.64]{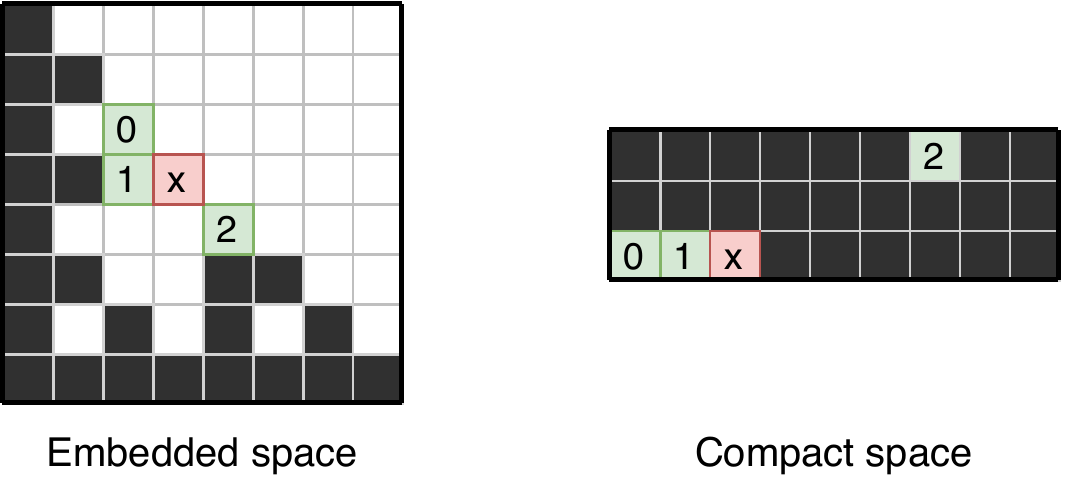}
\caption{A comparison of the neighborhood between expanded and compact space for a cell in the Sierpinski Triangle.}
\label{fig:neighbor}
\end{figure}

The challenge is to access the data-parallel neighborhoods efficiently, as these will no longer be Moore or Von-Neumann for every cell. Figure \ref{fig:neighbor} illustrates this difference (the placement of neighbors in compact space is properly explained in Section \ref{sec:proposal}).

This work presents Squeeze, a tensor core GPU approach capable of doing parallel computations, including stencil-like or nearest-neighbors simulations, on compact NBB fractals. It enables applications such as PDE solvers, cellular-automata, spin-model simulations, among others, to do efficient fractal simulation in compact space, as they rely on accessing neighboring cells to simulate the corresponding phenomena. 
Squeeze is based on the combination of two block-space maps that act between compact and embedded space, both adapted as tensor core MMA operations to further increase GPU performance. Experimental results show that Squeeze is up to $12\times$ faster and $315\times$ more memory efficient than a GPU-based bounding box approach, and can even match the performance of another recent GPU-based work that only improved performance but still sacrificed memory. 

The remaining sections of the manuscript cover related work (Section~\ref{sec:related-work}), the proposed Squeeze approach (Section~\ref{sec:proposal}), experimental results (Section~\ref{sec:results}) and conclusions (Section~\ref{sec:conclusions}).

\section{Related Work}
\label{sec:related-work}
This section will cover the first related work \cite{NAVARRO2020158, 8291959} in more detail, as it establishes the foundations for the present work. The rest of the works are synthesized and grouped by type of contribution.

\begin{figure}[ht!]
\centering
\includegraphics[scale=0.34]{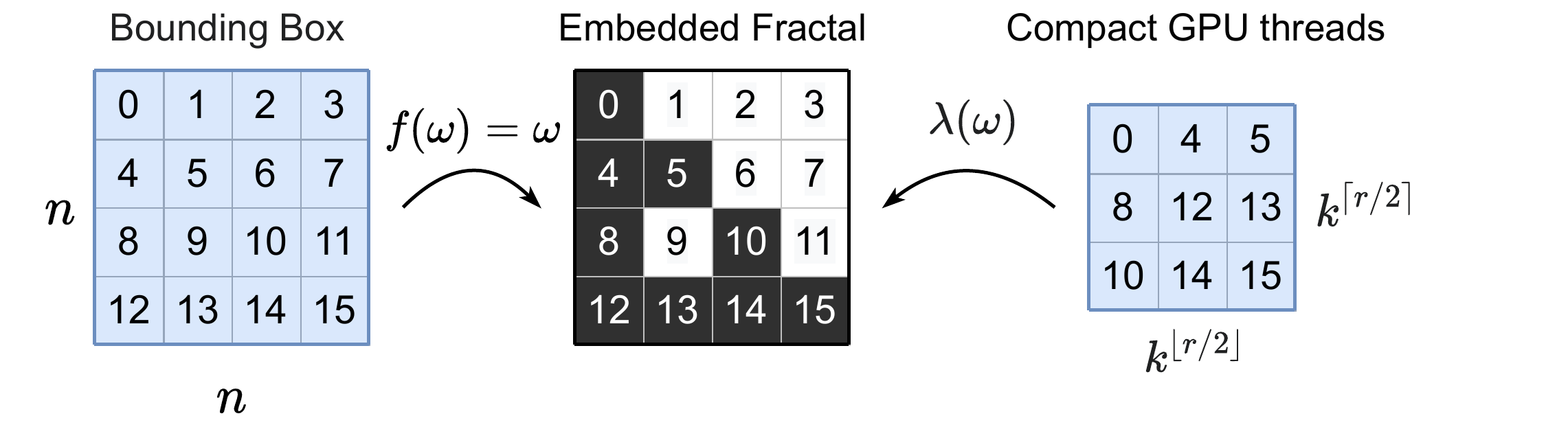}
\caption{Mapping of threads with BB and $\lambda(\omega)$ \cite{NAVARRO2020158}.}
\label{fig:bb-vs-lambda}
\end{figure}

Navarro \textit{et al.}~\cite{NAVARRO2020158, 8291959} proposed an efficient GPU Tensor-core accelerated thread map for NBB fractals, denoted $\lambda(\omega)$, where $\omega$ is a 2D coordinate. The authors report up to $9\times$ of speedup over a bounding-box approach, and up to an extra $40\%$ of speedup by using GPU tensor cores. The $\lambda(\omega)$ map allows using a reduced number of threads to reach the fractal data elements in an expanded embedded representation. In terms of CUDA programming, the approach proposed by Navarro \textit{et al.} compacts the CUDA Grid of thread-blocks to the minimum necessary to efficiently process the fractal, leading to a significant speedup. Figure \ref{fig:bb-vs-lambda} illustrates the benefits of using $\lambda(\omega)$ compared to bounding-box (BB). 

Although $\lambda(\omega)$ can solve problem P1 and improve performance significantly, it cannot manage all\footnote{The only case where $\lambda(\omega)$ can work in compact space is when threads operate just their own cell.} types of simulations in compact space, such as stencil or general nearest-neighbors ones. Therefore, $\lambda(\omega)$ cannot solve problem P2 properly, leaving an open research problem on finding an efficient GPU solution for P1 and P2.

The rest of the related work is grouped into three main topics: i) GPU processing in complex domains, ii) Compact GPU processing on sparse data iii) acceleration of non-AI tasks via GPU tensor cores.

\subsection{GPU processing in complex domains.}
 
Jung \textit{et al.}~\cite{Jung2008} developed an algorithm to map triangular (2-simplex) shaped data to a rectangular box to accelerate LU and Cholesky decomposition. The total memory used is reduced in half.

Ries \textit{et al.}~\cite{Ries:2009:TMI:1654059.1654069} developed a new method to compute the inverse of triangular matrices by developing a recursive parallel space mapping from a compact rectangular domain using GPU. The map's complexity is $O(\log_2(n))$.

Navarro \textit{et al.}~\cite{DBLP:conf/hpcc/NavarroH14,CLEI-2016-navarro,navarro2018competitiveness}
proposed a GPU block-space mapping for 2 and 3-simplex domains based on a linear numbering of discrete elements. Authors report an empirical speedup of up to $1.5\times$ and $2.3\times$ over a bounding-box approach, for 2 and 3-simplices, respectively.

\subsection{Compact GPU processing on sparse data.}

Zachariadis \textit{et al.}~\cite{ZACHARIADIS2020106848} proposed tSparse, an algorithm to accelerate sparse General Matrix Multiplication (GEMM) using Tensor Cores. Their approach use specific data structures and Look-up tables to efficiently save data.
This approach significantly reduces computational resources requirements. It is on average $1.53\times$ faster than other sparse GEMM techniques and the use of tensor cores provides an additional $68\%$ of extra performance.

Weber \textit{et al.}~\cite{WebBenSchStoFel13} presented a novel approach that significantly accelerates sparse matrix and vector multiplication in GPU. 
Authors report up to $13\times$ speedup.

Ferrando \textit{et al.}~\cite{FERRANDO2011628,8430608}
accelerated the processing of 3D cellular automaton simulations using an Octree to subdivide the 3D space. Their approach has a mixed GPU-CPU implementation with raw processing done in GPU and Octree update in CPU. 

\subsection{Acceleration of non-AI tasks via Tensor Cores.}

Carrasco \textit{et al.}~\cite{inproceedingsTC} studied the theoretical benefit of tensor core based arithmetic reductions. Authors conclude that a tensor core based reduction is, indeed, faster. Continuing with their work \cite{9147055}, an implementation of said techniques resulted with a empirical speedup of $3.2\times$ over traditional CUDA cores in large problem sizes. 

Dakkak \textit{et al.}~\cite{10.1145/3330345.3331057} used tensor core to accelerate scan and arithmetic reductions in GPU achieving up to $100\times$ speedup for small sized problems in reduction and up to $3\times$ for scan. 

The main difference of 
our proposal with respect to the
described related work, 
is the focus on handling discrete fractal domains in compact space, and also the extra acceleration by adapting all map computations to GPU tensor cores. The next Section explains Squeeze; the proposed approach that allows efficient GPU tensor core processing on compact NBB fractals.

\section{Overview of Squeeze}
\label{sec:proposal}
The proposed approach is a GPU scheme capable of processing any NBB fractal in compact space. As a result, it increases GPU performance and reduces memory usage when compared to an expanded bounding-box approach.
Squeeze combines two GPU thread maps: $\lambda(\omega)$, an existing state-of-the~art map \cite{NAVARRO2020158} that transforms from compact space to expanded space, and $\nu(\omega)$, a new proposed map that transforms from expanded space to compact space. Using the two maps in conjunction, it is possible to do any kind of discrete simulation using only the compact space.

The notation $\fracnot{n}{k}{s}$ will be used to denote a fractal in the NBB class and to introduce the values $n, k, s$.  Here, $n \in \mathbb{N}$ is the linear size of the fractal along one axis, $k \in \mathbb{N}$ the number of self-similar replicas generated by its transition function and $s \in \mathbb{N}$ the growth ratio of $n$ in the next scale level, along an axis. 
For example, the Sierpiński Carpet (Figure \ref{fig:carpet-fractal-embedded}) is $\fracnot{n}{8}{3}$ and the empty bottles fractal (Figure \ref{fig:nbb-construction}) is $\fracnot{n}{7}{3}$. 
Parameters $k$ and $s$ are specific to a fractal and $n$ scales up by factors of $s$ as the fractal level increases. The space used by a fractal, denoted as $\mathcal{V}(\fracnot{n}{k}{s})$ may be expressed as:
\begin{equation}
    \mathcal{V}(\fracnot{n}{k}{s}) = k^r
\end{equation}
where $r = \log_{s}(n)$ is defined as the scale level. 
Many different NBB fractals can be described using the same parameters. A table of NBB examples can be found in the work that formulated $\lambda(\omega)$ \cite{NAVARRO2020158}.
 
The presentation of Squeeze continues with a explanation of how any NBB fractal can be compacted into a rectangular region, then with a general view of how neighborhood exploration is achieved in compact space, which includes a brief revisit to $\lambda(\omega)$, and a detailed formulation of the new map $\nu(\omega)$ and its adaptation to GPU tensor cores.

\subsection{Compacting NBB Fractals}

Any NBB fractal expressed in the expanded embedded bounding-box representation, denoted $D^2$, has a corresponding compact representation, denoted $D^2_c$. This compact representation is a rectangular region of size $k^{\lfloor \frac{r}{2} \rfloor}\times k^{\lceil \frac{r}{2} \rceil}$ elements, which satisfies $\mathcal{V}(\fracnot{n}{k}{s}) = k^r$. 

\begin{figure}[ht!]
\centering
\includegraphics[scale=0.21]{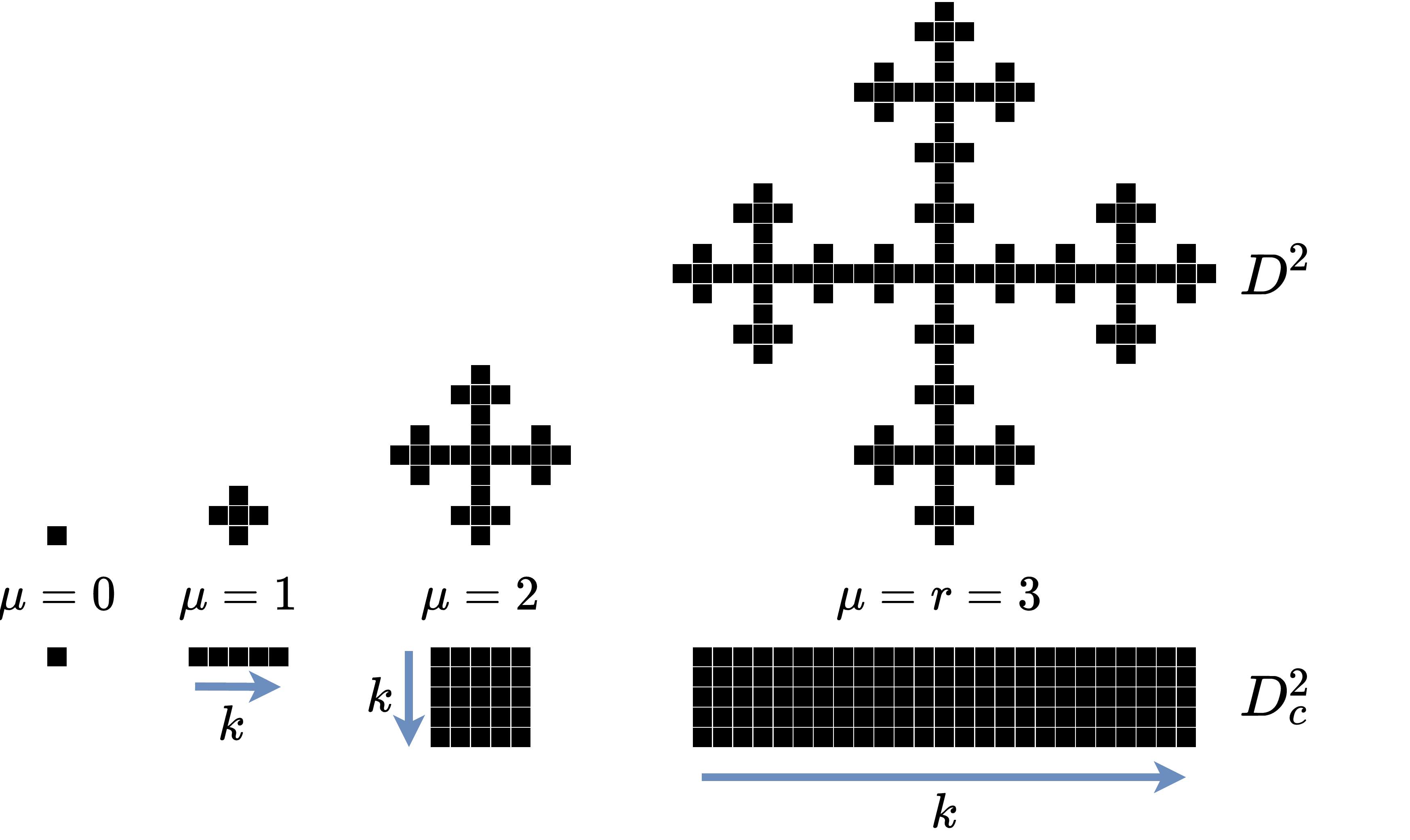}
\caption{Compact space logic for the Viczek fractal $\fracnot{27}{5}{3}, r=3$.}
\label{fig:nbb-construction-viczek}
\end{figure}

Given a NBB fractal $\fracnot{n}{k}{s}$ in expanded embedded space, with $r = \log_{s}(n)$ the number of levels it has, the packing of its discrete data-elements into compact space follows a logic similar to an \textit{unrolling} process of $r$ steps. Let $\mu=0$ refer to $\fracnot{n}{k}{s}$ at level $0$; at this level the compact space has only one data element and coincides with the embedded space as well. At $\mu=1$ (odd), the compact space is scaled up in $x$ by $k$ times (replication factor). At $\mu=2$ (even), the compact space is scaled up in $y$ by $k$. This process continues the same way for odd and even values until $\mu = r$. 
Figure \ref{fig:nbb-construction-viczek} shows as an example the compaction logic for the Viczek fractal.



\subsection{Exploring Compact Fractal Space}
Parallel exploration in a fractal's compact space is a special procedure, because a cell's neighborhood in $D^2$ in drastically different when seen in $D_c^2$.
Squeeze combines the already known $\lambda(\omega)$ map, with a the new map $\nu(\omega)$ (formulated ahead in Section \ref{subsec:formulation-nu}) in such a way that the former acts as a tensor core-accelerated function from compact-space to embedded-space, while the latter acts as a tensor-core-accelerated function from embedded-space to compact-space. By using the two maps, it is possible for all fractal locations to explore their neighborhoods without expanding the fractal into embedded space in memory, thus preventing memory sacrifices. 

\begin{figure}[ht!]
\centering
\includegraphics[scale=0.58]{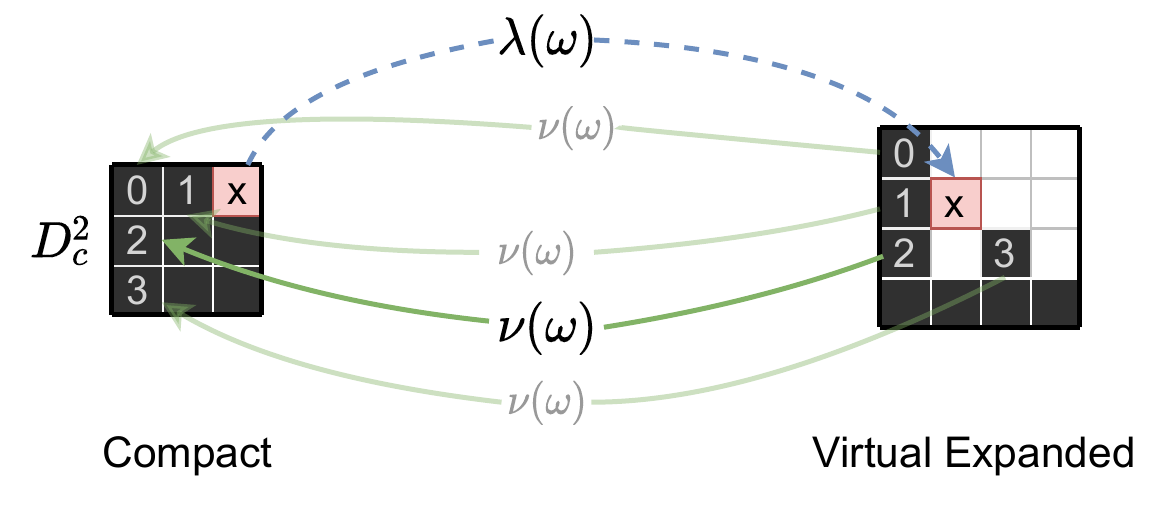}
\caption{Illustration of the process to access neighboring cells in a Sierpinski Triangle. The red cell represents the simulated element.}
\label{fig:example-nu}
\end{figure}

To explore fractal neighborhood in compact space, Squeeze first uses $\lambda(\omega)$ to transform the thread's data-element location into expanded embedded space. Once the thread's location is transformed, it can easily offset its coordinate to define a neighborhood in this virtual expanded space which is transitory and does not use GPU memory. Once a neighborhood is defined, each coordinate is transformed back to compact space with $\nu(\omega)$. With this, each neighbor is now identified in compact space and GPU threads can access the element in memory and continue the application's computation transparently. This approach performs at most one execution of $\lambda(\omega)$ map and $\ell$ executions of $\nu(\omega)$, where $\ell$ is the number of neighbors to explore.
Figure \ref{fig:example-nu} illustrates how the Squeeze approach uses both maps in conjunction.


\subsection{Revisiting $\lambda(\omega)$}
Although $\lambda(\omega)$ is extensively described in \cite{NAVARRO2020158}, we provide a short summary that will be useful to formulate the new map $\nu(\omega)$. Map $\lambda(\omega)$ is formally defined as a thread-block mapping that transforms a coordinate from compact parallel space to a unique coordinate in the embedded space. This mapping can be done in $\mathcal{O}(\log_2(\log_s(n))$ time. 

\begin{figure}[ht!]
\centering
\includegraphics[scale=0.38]{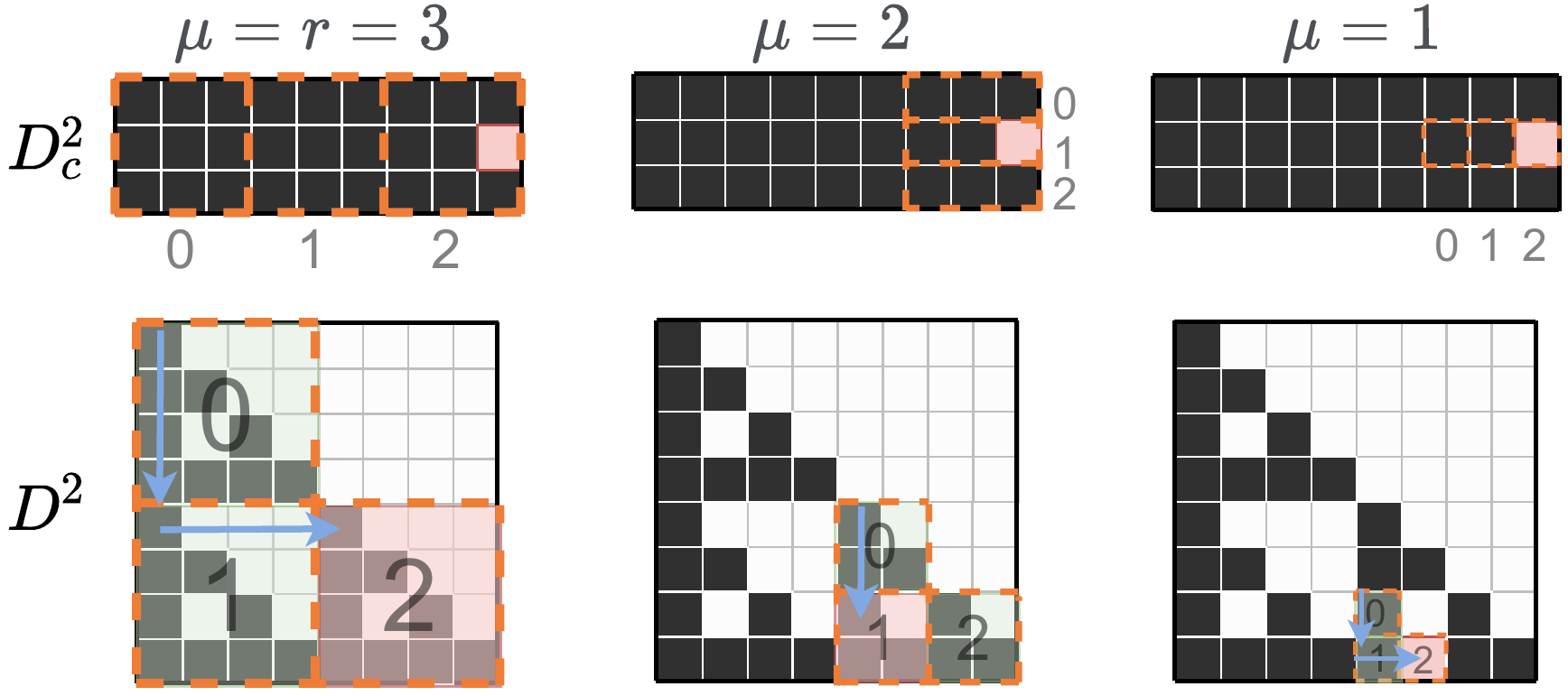}
\caption{The action of $\lambda(\omega)$ illustrated for a single element in a Sierpinski Triangle at level $r=3$. The red element is mapped from compact space to expanded embedded. The dashed orange blocks represent the scope of the current $\mu$ where the replica is identified, and the red shaded region is the current known sub-region where the cell is guaranteed to belong. The blue arrows are the offsets to accumulate at that iteration.}
\label{fig:lambda-example}
\end{figure}

The basic intuition behind $\lambda(\omega)$ is a coordinate offset accumulation in the compact space by detecting in which replica does the coordinate $\omega$ resides and offset accordingly in relation to its value. This is repeated in a top-down manner passing through the different scale levels, \textit{i.e.}, from $\mu=r$ to $u=1$ (level $u=0$ does not generate any offset). This folding scheme is illustrated in Figure \ref{fig:lambda-example}.  

The map $\lambda(\omega): \mathbb{N}^2 \mapsto \mathbb{N}^2$ is defined as follows 
\begin{align}
    \lambda(\omega)   &= \sum_{\mu=1}^{r} \Delta_{\mu}
\end{align}
where $\omega$ is a parallel resource coordinate (such as a thread or a block of threads). The process is a summation of offsets $\Delta_\mu$ at each level, which are defined as 
\begin{align} 
    \Delta_\mu &= \tau(\beta_\mu)\cdot s^{\mu-1}
\end{align}
where $\tau$ is an normalized offset in the number of replicas along each axis (the scaling value $s$). In practice, $\tau$ acts through $H_\lambda[\beta_\mu]: \mathbb{N} \mapsto \mathbb{N}^2$ which can be a look-up table of size $k$ that returns 2D coordinates, or a direct arithmetic hash if the replica patterns allow it. With this, $\tau$ is defined as
\begin{align}
    \tau(\beta_\mu) = H_\lambda[\beta_\mu] = (\tau_x, \tau_y),\ \  \tau_x, \tau_y \in [0..s-1]
\end{align}
For example, in the Sierpinski triangle we have that $s=2$ and the number of replicas is $k=3$. Therefore, the possible outcomes for $\tau$ would be $\tau(0) = (0,0)$, $\tau(1) = (0,1)$ and $\tau(2) = (1,1)$ to refer the top, middle and right replicas, respectively. The proper scaling of the offset is done by the $s^{\mu-1}$ factor. Input $\beta_\mu$ is an auxiliary index defined as
\small
\begin{equation}
        \beta_\mu(\omega) = \Big( \frac{\omega_x(\mu \ \mathbf{mod}\ 2) + \omega_y((\mu+1 )\ \mathbf{mod}\ 2)}{k^{\lceil \frac{\mu}{2} \rceil-1}}\Big) \ \mathbf{mod}\ k%
\end{equation}
\normalsize
and has a range of $[0, k-1]$. It identifies, within scale level $\mu \in [1..r]$, which of the $k$ replicas of the fractal does the coordinate $\omega$ belong to, 

To summarize $\lambda(\omega)$, this map serves as a mechanism to go from compact space to expanded embedded space. In the next sub-section, we formulate $\nu(\omega)$, the map that allows going from expanded embedded space back to compact space.

\subsection{Formulation of $\nu(\omega)$}
\label{subsec:formulation-nu}
In order for Squeeze to work, we require a transformation from expanded space to compact space, \textit{i.e.}, $\mathbf{D}^2 \mapsto \mathbf{D}_c^2$. We propose $\nu(\omega): \mathbb{N}^2 \mapsto \mathbb{N}^2$, a map that acts as the inverse of $\lambda(\omega)$. As with $\lambda(\omega)$, $\nu(\omega)$ assumes that the origin $(0,0)$ is located at the upper-left corner of both $\mathbf{D}^2$ and $\mathbf{D}^2_c$ spaces, and the $x,y$ axes increase to the right and downwards, respectively.  

\begin{figure}[ht!]
\centering
    \includegraphics[scale=0.4]{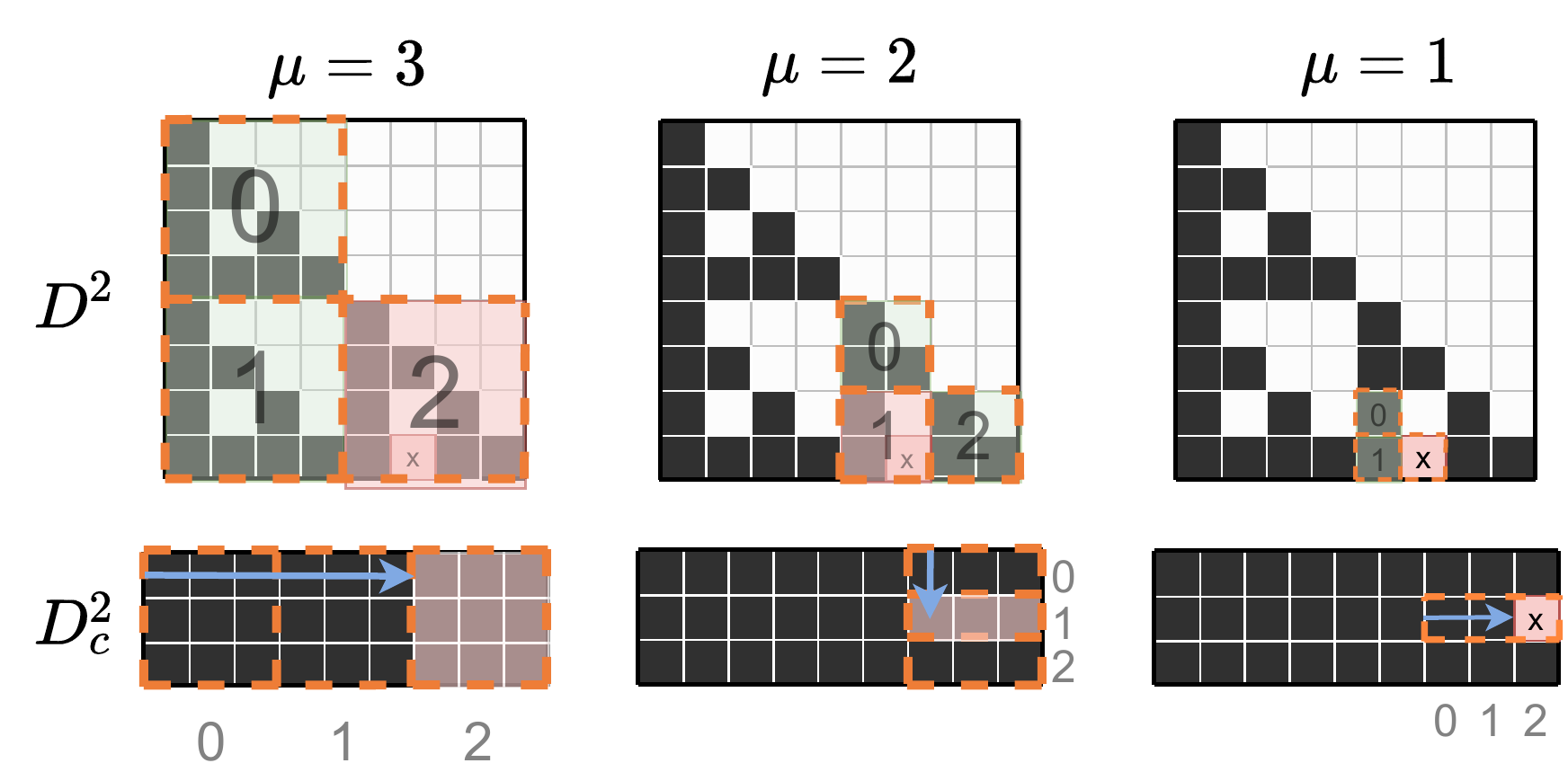}
\caption{Example of the process done by $\nu(\omega)$ with a Sierpinski Triangle of $r=3$. The red element is mapped from embedded space to compact space. The dashed region is the actual scope of iteration $\mu$ where the replica is identified. The blue arrows are the offsets to accumulate at each iteration.}
\label{fig:nu}
\end{figure}

To formulate $\nu(\omega)$, the offset accumulation approach is used again, but now checking on the expanded space what replica does element $\omega$ belongs to at each level using a top down scheme, and applying the offset into the compact space starting from $(0,0)$ (upper-left). Going through all the scale levels, $\mu=r$ to $\mu=1$ ($\mu=0$ can be skipped as it gives zero offsets), allows collecting all the offset contributions. For even values of $\mu$, the offset is accumulated in the $x$ axis and for odd values, in the $y$ axis. The magnitude of the offset depends on the actual value of $\mu$. Figure \ref{fig:nu} illustrates the process in a Sierpinski Triangle of $r=3$.

The formulation of $\nu(\omega)$ follows a similar procedure that of $\lambda(\omega)$ \cite{NAVARRO2020158}. Let $H_\nu(\theta_\mu): \mathbb{N}^2 \mapsto \mathbb{N}$ be a look-up table of size $k$, that given a coordinate $\theta_\mu = (\theta_{\mu,x}, \theta_{\mu,y})$ returns a value in the range $[0, k-1]$ that represents which replica does the coordinate $\omega$ belong to at level $\mu$ in the embedded fractal. The coordinate $\theta_\mu$ is defined as:
\begin{align} \label{eq:theta}
\theta_{\mu, x|y} &= \left\lfloor \frac{\omega_{x|y}\ \mathbf{mod}\ s^{\mu}}{s^\mu} \right \rfloor
\end{align}


Let $\Delta_\mu^\nu$ (with superscript $\nu$ to denote that it refers to $\nu(\omega)$) denote the offset of the replica at a particular scale level. $\Delta_\mu^\nu$ is defined as:

\begin{equation} \label{eq:t}
\Delta_\mu^\nu = k^{\left\lfloor\frac{\mu-1}{2}\right\rfloor} 
\end{equation}
for $\omega_x$ and $\omega_y$ coordinates. Let $f(\mu)$ denote a filter function that alternates the accumulation of offsets between axes $x$ and $y$ as $\mu$ increases,
\begin{align} \label{eq:f}
    f(\mu) &= \left(f_x(\mu), f_y(\mu)\right)\\
    f_x(\mu) &= (\mu-1)\ \mathbf{mod}\ 2\\
    f_y(\mu) &= (\mu)\ \mathbf{mod}\ 2
\end{align}
Combining the offsets with the lookup-table results and the filters, $\nu(\omega)$ becomes:
\begin{align}
\nu(\omega) &= (\nu_x(\omega), \nu_y(\omega)) \\
\label{eq:nux}
\nu_x(\omega) &= \sum_{\mu=1}^{r}\Delta_\mu^\nu \cdot H_{\nu}[\theta_{\mu}]\cdot f_x(\mu) \\
\label{eq:nuy}
\nu_y(\omega) &= \sum_{\mu=1}^{r}\Delta_\mu^\nu \cdot H_{\nu}[\theta_{\mu}]\cdot f_y(\mu).
\end{align}
The map $\nu(\omega)$ can be computed by each GPU thread in parallel. However, in practice a better practice is to map blocks of threads in order to allow thread collaboration in the computation of $\nu(\omega)$ and data-locality at a small scale.

\subsection{Moving from Thread-level to Block-level}
\begin{figure}[ht!]
\centering
\includegraphics[scale=0.35]{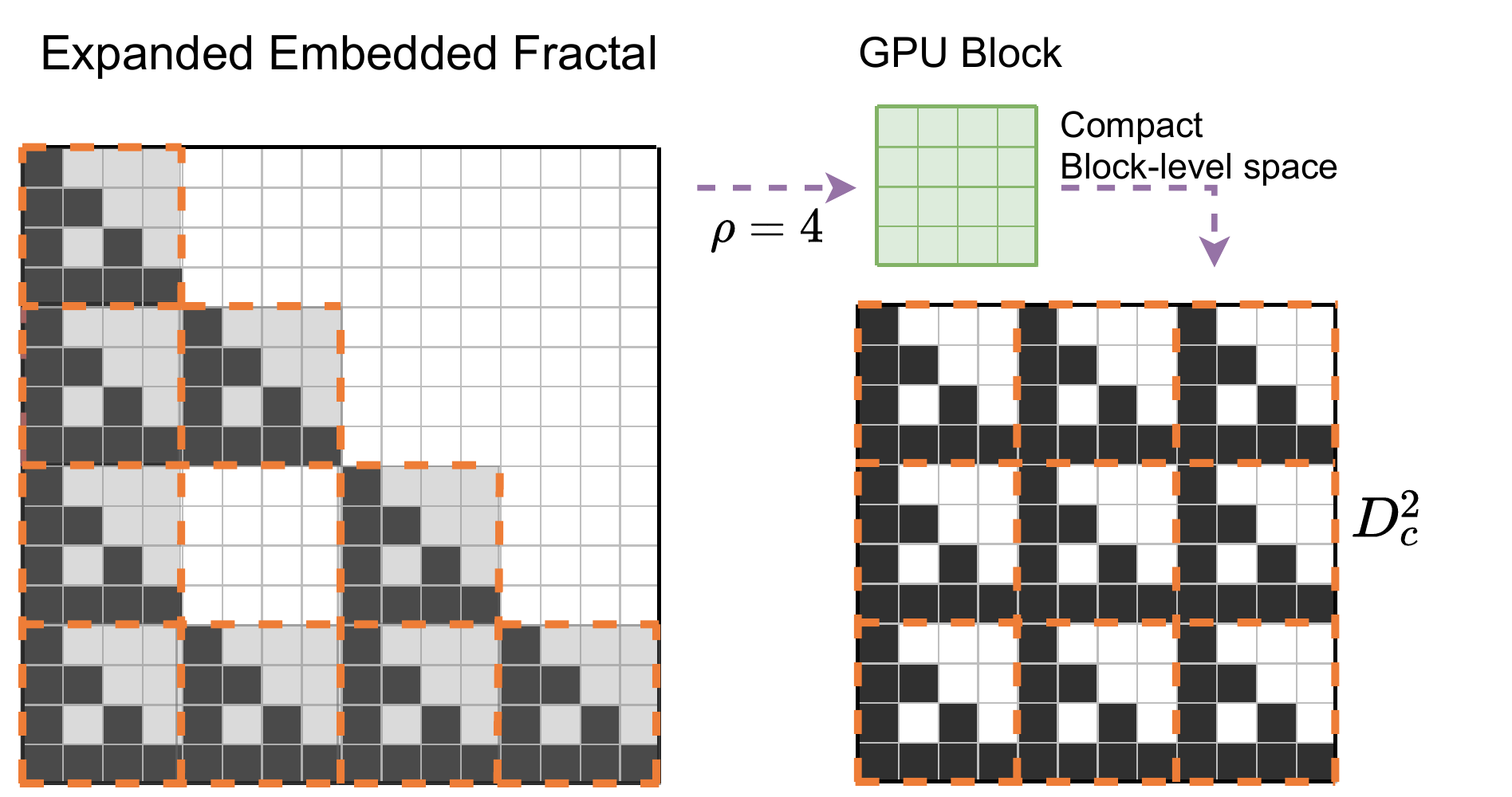}
\caption{Block-level Squeeze. In this example, each block has $4\times 4$ elements, making the Sierpinski Triangle, of level of $r=4$, become a coarser one of $r_b=2$.}
\label{fig:coarser}
\end{figure}

To make the process more efficient on GPU, one can apply Squeeze at a block-level, that is, instead of mapping the thread coordinates, to map the block-coordinates. With this change, all $\rho \times \rho$ threads of a block now represent one coarse coordinate. In other words, block-level Squeeze can be seen as handling a lower level version of the fractal, where now $r,n$ change to $r_b=r-\log_2(\rho)$, and $n_b = \frac{n}{\rho}$. This has the benefit of requiring less operations, allowing thread cooperation and producing memory locality within each block. Thread cooperation in a block allows doing a parallel reduction on Eqs. (\ref{eq:nux}) and (\ref{eq:nuy}), resulting in a running time of $O(\log_2\log_s(n))$ for $\nu(\omega)$. It is also worth noticing that in block-level Squeeze, space will be compacted at block-level, as expected, and inside each block one would find a small, constant size, expanded embedded fractal. Similar to quick-sort or other divide and conquer algorithms, applying micro-brute-force solutions at the last levels of a fractal structure can prove to be more efficient in terms of parallelism and locality. Figure \ref{fig:coarser} illustrates a block-level compact space. 

These micro-fractals indeed introduce an extra memory usage and may have some unused threads, but this overhead is constant as long as the size of the micro-fractals is in terms of the block-size which is constant.  


Although block-level Squeeze has an efficient upper bound of $\mathcal{O}(\log_2(\log_s(n)))$ for both $\nu(\omega)$ and $\lambda(\omega)$, this performance can be further accelerated with tensor core units (TCU). 

\subsection{Accelerating Squeeze with Tensor Core Units}
Today modern GPUs have up to hundreds of Tensor Core Units (TCUs). Each TCU provides a hardware-level matrix-multiply-accumulate (MMA) operation defined as
\begin{equation}
    D = A\times B + C
\end{equation}
that when called, is executed by a warp of threads and runs in parallel with the rest of the tensor core units of the GPU chip. Matrices $A,B,C,D$ can have  in the order of $256$ elements, or more if precision is relaxed to FP16 or less.  

The acceleration of Squeeze with tensor cores is achieved by encoding the sum of products found both in $\lambda(\omega)$ and $\nu(\omega)$. The Tensor Core adaptation of $\lambda(\omega)$ was already done in its previous work \cite{NAVARRO2020158}, therefore this subsection describes the process for the remaining map $\nu(\omega)$. The adaptation consists of encoding the sum of products found in Eq.~(\ref{eq:nux}) and Eq.~ (\ref{eq:nuy}) as MMA operations. The encoding is 
\small
\begin{align}
        \label{eq:mma-a}
        A &=
        \begin{pmatrix}
        \Delta_1^\nu f_x(1) & \Delta_2^\nu f_x(2) & \dots & \Delta_{r}^\nu f_x(r)\\
        \Delta_1^\nu f_y(1) & \Delta_2^\nu f_y(2) & \dots & \Delta_{r}^\nu f_y(r)\\
        0          & 0          & \dots & 0\\
        \vdots & \vdots & \ddots & \vdots\\
        0 & 0 & \dots & 0\\
        \end{pmatrix}\\
        \label{eq:mma-b}
        B &=
        \begin{pmatrix}
        H_\nu[\theta_1] & 0 & \dots & 0\\
        H_\nu[\theta_2] & 0 & \dots & 0\\
        \vdots & \vdots & \vdots & \ddots & \vdots\\
        H_\nu[\theta_{r}] & 0 & \dots & 0
        \end{pmatrix}\\
\end{align}
\normalsize
with $C$ being a zero-matrix. These MMA operations are assigned one per warp, and given that the MMA can be perceived as a $O(1)$ operation (hardware level), potential performance speedup is expected. The programming of tensor cores was done using CUDA's  WMMA API available. Current restrictions on CUDA tensor cores specify that each tensor core matrix (known as fragment in CUDA) can have at most $256$ elements when using FP16 multiplication ($A \times B$) and FP32 accumulation ($+C$). This work chose $16 \times 16$ matrices for Squeeze, as the F16/FP32 configuration provided correct results.   

\subsection{Theoretical Memory Reduction Factor}
\begin{figure}[ht!]
\centering
\includegraphics[scale=0.44]{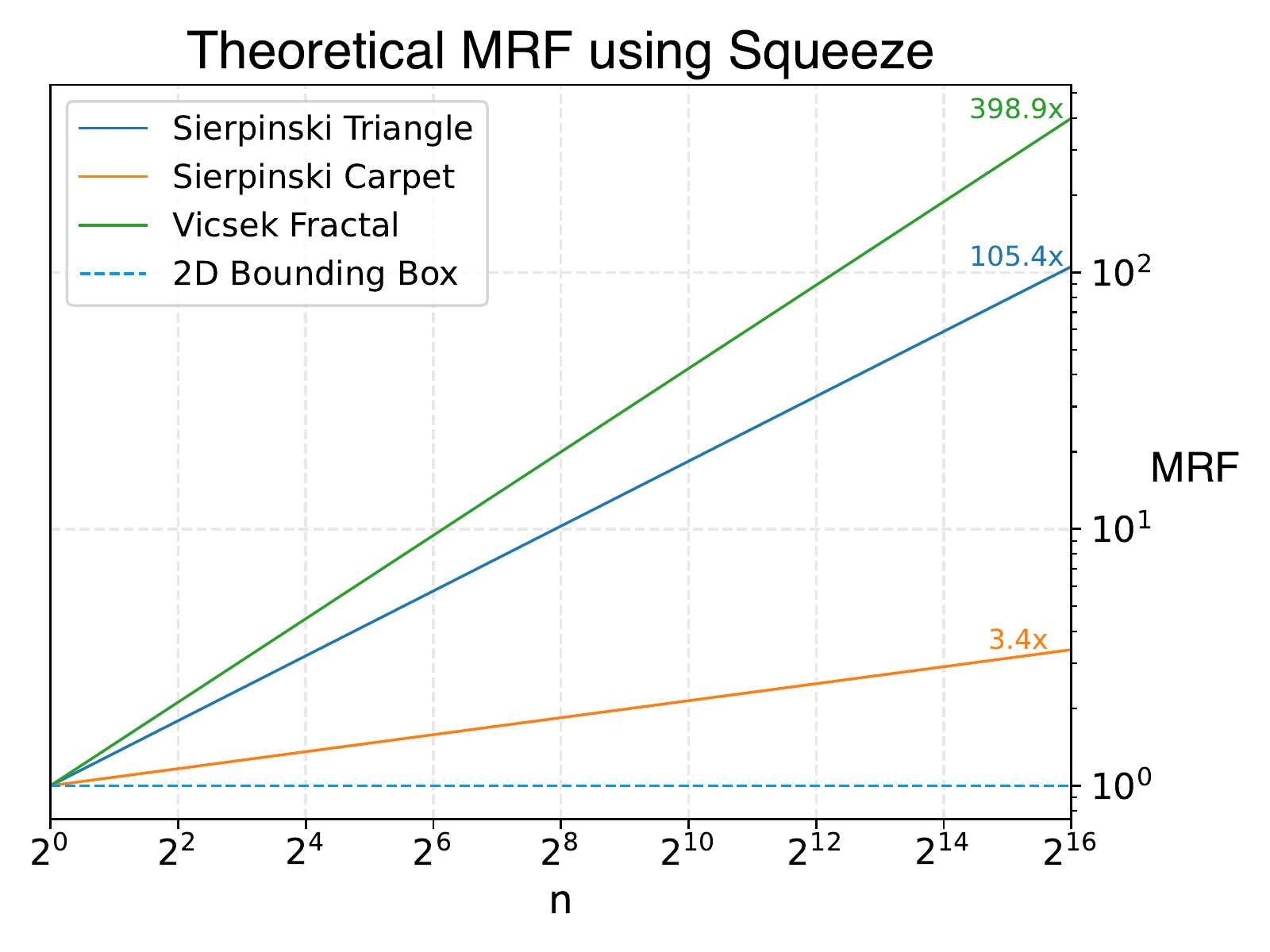}
\caption{Theoretical memory-reduction-factor of Squeeze.}
\label{plot:theoretical}
\end{figure}

Figure \ref{plot:theoretical} shows the theoretical Memory Reduction Factor (MRF) of Squeeze over a expanded bounding-box (BB) scheme, for three NBB fractals.

From the plot it is possible to note that the MRF grows exponentially as the size of the fractal increases. At $n=2^{16}$, Squeeze obtains an MRF close to $400\times, 105\times$ and $3.4\times$ for the Viczek, Sierpinski Triangle and Sierpinski Carpet fractals, respectively. This enables the possibility to process in a single GPU fractals that did not fit before.

\section{Experimental Setup and Results}
\label{sec:results}
\begin{figure}[ht!]
\centering
\includegraphics[scale=0.20]{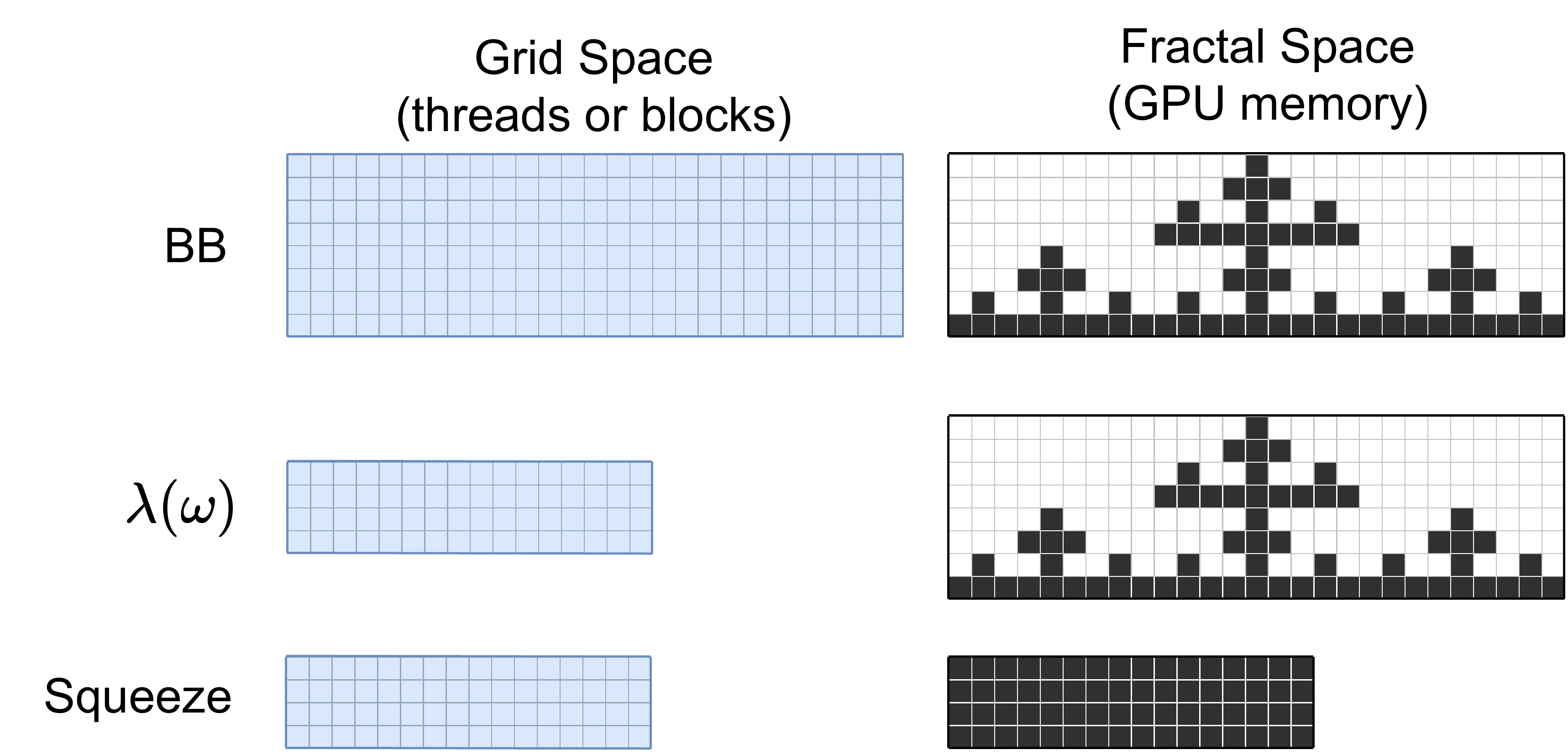}
\caption{Comparison of grid and memory spaces of the three approaches. The chosen fractal is named chandelier.}
\label{fig:comparison}
\end{figure}

Experimental tests were employed using Conway's game of life running on a Sierpinski Triangle as a case study,  considering a Moore's neighborhood in expanded space. Only elements that belong to the fractal are simulated as well as considered as neighbors for the others, \textit{i.e.}, the \textit{holes} were skipped. Life/Death conditions were adapted for this same reason. Three different GPU-based fractal processing approaches were implemented\footnote{The repository will be available in the published version.}:
\begin{enumerate}
    \item \textbf{BB}: Expanded grid and fractal representation in memory. It is the \textit{classic approach}.
    \item \textbf{$\bm{\lambda(\omega)}$}: Compact grid and expanded fractal \cite{NAVARRO2020158}. It is a \textit{state-of-the-art approach}. 
    \item \textbf{Squeeze}: Compact grid and compact fractal. It is the \textit{proposed approach}.
\end{enumerate}
Figure \ref{fig:comparison} shows a comparison of the grid and memory requirements for each approach, using as an example the \textit{chandelier} fractal. 

The performance metrics are the execution time, denoted $T$, and Speedup denoted $S$. Execution time was measured as the average time of $100$ runs of $1000$ simulation iterations for each hardware setup (listed in Table \ref{table_hardware}). This guaranteed an standard error lower than $1\%$. The Speedup is defined as
\begin{equation}
    S=\frac{T_{ref}}{T_{comp}}
\end{equation}
where $T_{ref}$ is reference time, \textit{e.g.} the bounding-box running time, and $T_{comp}$ is the time of the tested approach. Parameters $r$ and $\rho$ were tested in the range $r = [0, 20]$ and $\rho = [2^0..2^5]$, respectively. A shared and non-shared memory version of each method was tested. The results presented only show the fastest version of each one. For BB and $\lambda(\omega)$ the fastest one was without\footnote{As a side note, we observed that for stencil computations, classic \textit{shared-memory} practices that were known to produce the fastest implementations in older GPUs (such as Nvidia K40), may no longer produce such effect in newer GPU architectures. We suspect it may relate to the improvements on the L2 cache.} and Squeeze with shared memory due to the extra memory accesses on local space. Finally, both $\lambda(\omega)$ and $\nu(\omega)$ use tensor core acceleration.

The computer setups used are listed in Table \ref{table_hardware}. 
\begin{table}[ht!]
\caption{Hardware setups used.}
\begin{center}
\begin{tabular}{| c | l | l |}
\hline
  & HW	&	Model\\
\hline
     & GPU	    &	TITAN V, 5120 cores, 12GB \\
   A & TCU      &   640 Tensor Cores First Gen\\
     & CPU	    &	Intel i7-6950X 10cores\\
 \hline
     &  GPU     &	TITAN RTX, 4608 cores, 24GB\\
   B &  TCU     & 576 Tensor Cores Second Gen\\
     &  CPU	    &	Intel i7-6950X 10-core\\
\hline
     & GPU	&	A100, 6912 cores, 40GB\\
   C & TCU  &   432 Tensor Cores Third Gen\\
     & CPU	&	$2\times$ AMD Epyc 7742, 64cores\\
\hline
\end{tabular}
\end{center}
\label{table_hardware}
\end{table}

Setup C is a DGX A100 node of the Patag\'on Supercomputer \cite{patagon-uach} from \textit{Austral University of Chile}.


\subsection{Case study: Sierpinski Triangle}
Experimental tests were done on the Sierpinski Triangle as an NBB fractal case study, simulating John Conway's game of life adapted to fractals. The reason why this fractal was chosen is mostly because it is one of the most recognized ones and has resulted to be a key structure for several applications, such as antenna construction~\cite{855489,664115},
cellular automata simulations~\cite{Ohi2001,RevModPhys.55.601},
fractal molecular ensembles \cite{shang2015}, DNA self-replication~\cite{rothemund2004}, among others. In NBB notation, the Sierpiński triangle denoted as $\fracnot{n}{3}{2}$, with $k=3$ and $s=2$.

Replacing the Sierpinski fractal specific parameters in $\nu(\omega)$ yields\footnote{For the Sierpinski Triangle version of $\lambda(\omega)$, see Navarro \textit{et al.}~\cite{NAVARRO2020158}}:
\begin{align} \label{eq:siertx}
\Delta_\mu^\nu &= 3^{\frac{\mu-1}{2}}\\
\nu_x(\omega) &= \sum_{\mu=1}^{r} 3^{\frac{\mu-1}{2}} H_\nu[\theta_{\mu}] f_x(\mu)\\
\nu_y(\omega) &=  \sum_{\mu=1}^{r} 3^{\frac{\mu-1}{2}} H_\nu[\theta_{\mu}] f_y(\mu)
\end{align}

\noindent where $H_\nu(\theta_\mu): \mathbb{N}^2 \mapsto \mathbb{N}$ enumerates each replica as: $0$ at the top, $1$ the middle and $2$ the right one. In the particular case of Sierpinski triangle, it is possible to use the hash expression 
\begin{equation} \label{eq:sierh}
H_\nu[\theta_\mu] = \theta_{\mu, x} + \theta_{\mu, y}
\end{equation}  
which is equivalent to the look-up table $H_\nu[(0,0)] = 0$, $H_\nu[(0,1)] = 1$, $H_\nu[(1,1)] = 2$.

In the tested application, the cellular automaton simulation, each thread accesses a fractal adaptation of Moore's neighborhood, that means at most eight $\nu(\omega)$ maps for the neighbors that needs to be brought back to the compact space. This neighborhood upper bound of eight neighbors brings the possibility to group up to eight $\nu(\omega)$ maps into one tensor core MMA operation, because the matrices are of size $16 \times 16$. Also, for technical simplicity the Tensor core approach was implemented for block sizes $\rho=16$ and $\rho=32$, other block-sizes use a regular CUDA computation of $\lambda(\omega)$ and $\nu(\omega)$.

\subsection{Performance Plots}
\begin{figure*}[ht!]
\begin{center}
\includegraphics[scale=0.30]{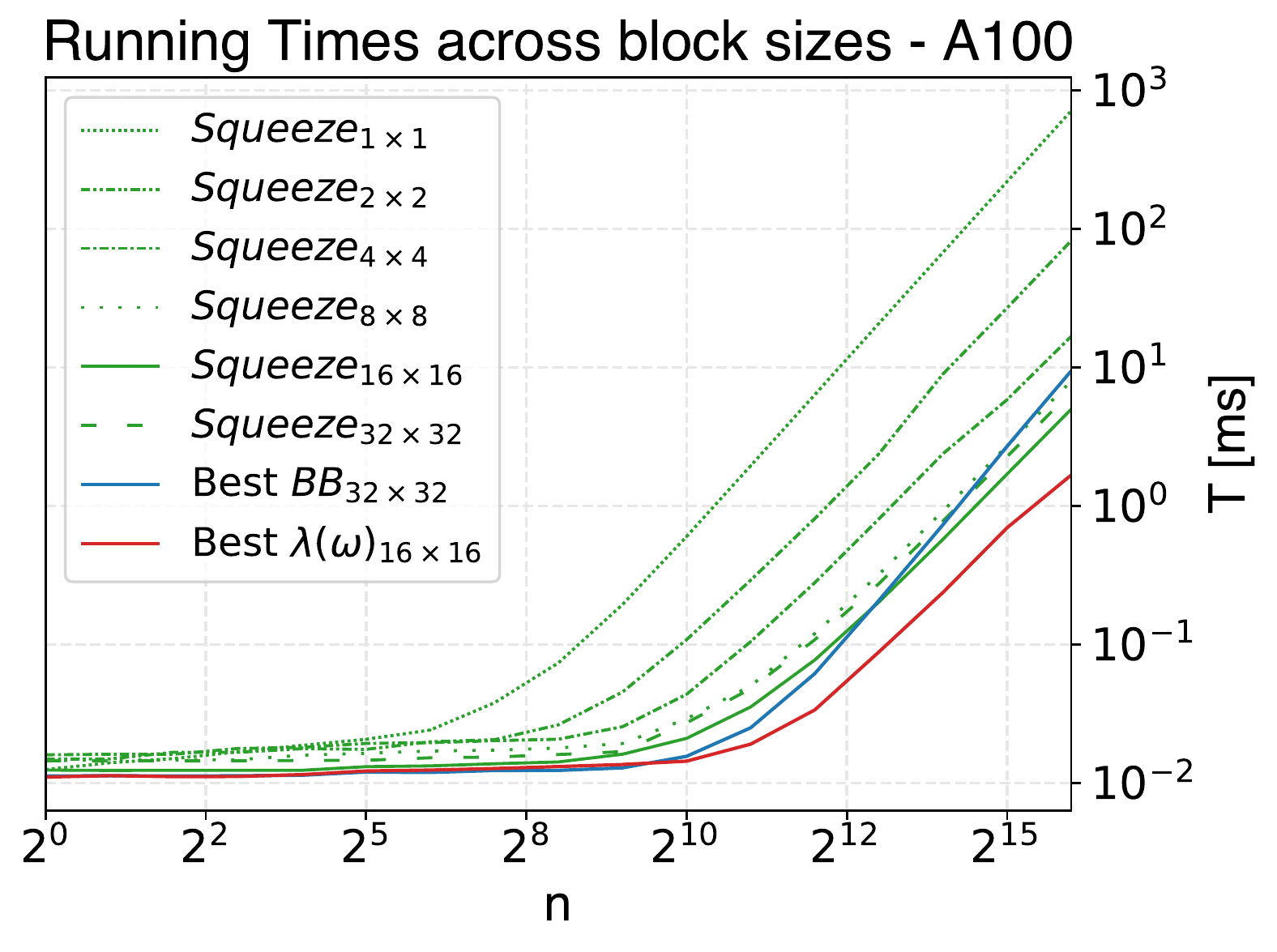}
\includegraphics[scale=0.30]{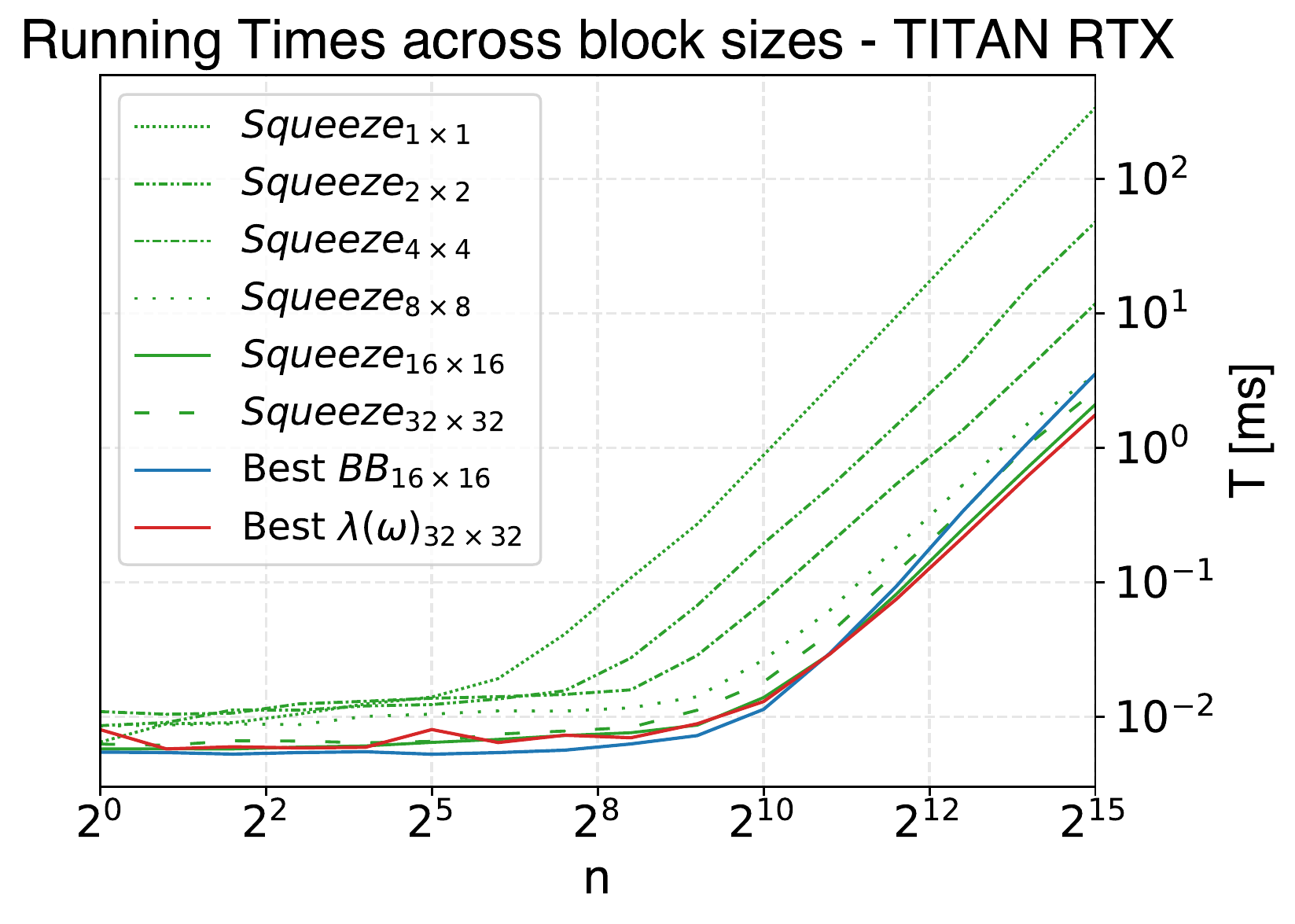}
\includegraphics[scale=0.30]{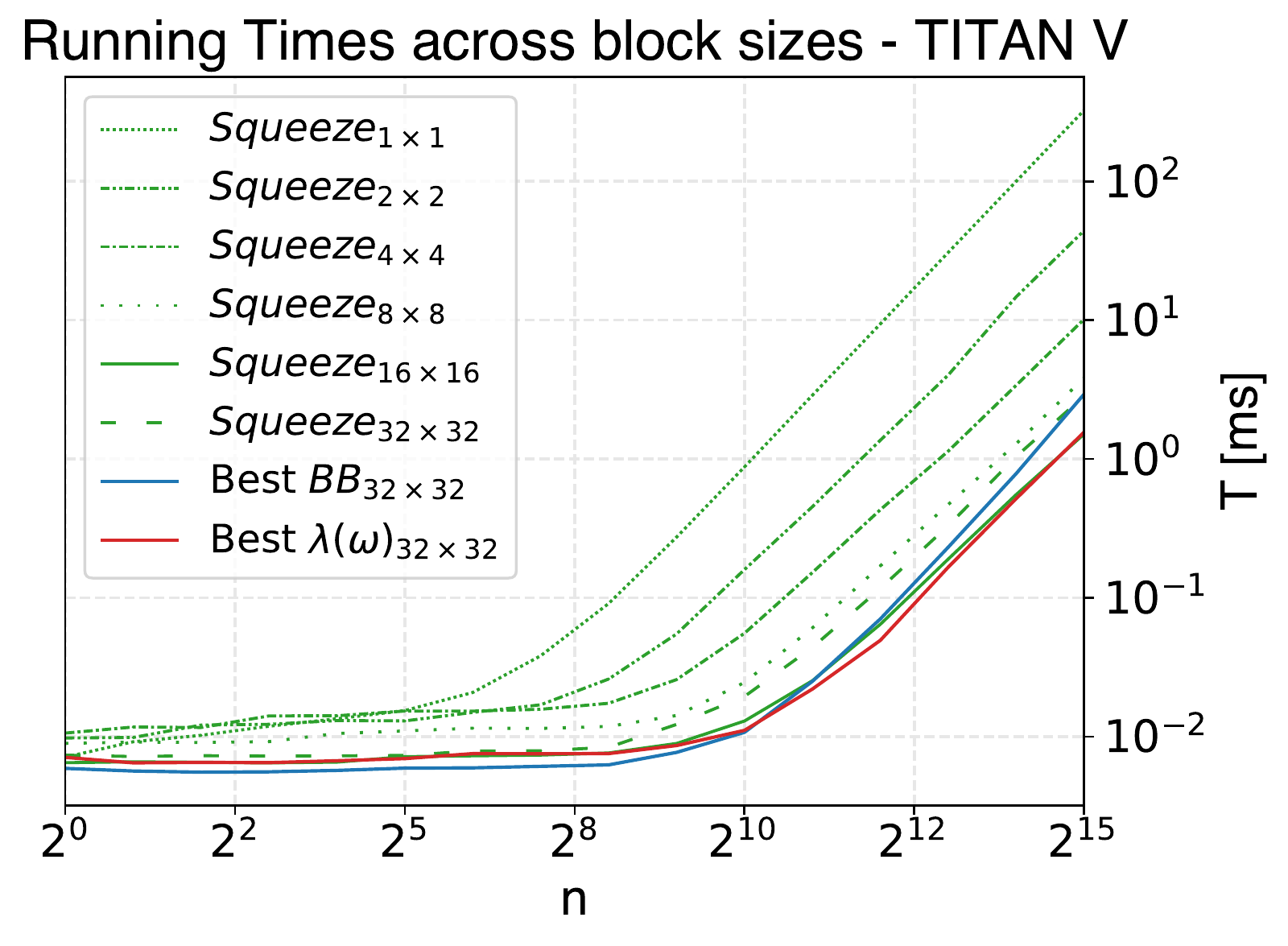}

\caption{Execution times of the 3 approaches; BB, $\lambda(\omega)$ and Squeeze.}
\label{fig:execution-time}
\end{center}
\end{figure*}

Figure \ref{fig:execution-time} presents Squeeze's execution time $T$ at different block sizes $\rho$ next to the fastest versions of BB and $\lambda(\omega)$ in terms of chosen block size.

Starting with the NVIDIA A100,  Once $n>2^{12}$, Squeeze's $\rho=16$ and $32$ configurations become significantly faster than BB (blue curve). On the other hand, for smaller problems in the range $n \leq 2^{12}$, all configurations of Squeeze ran slower than the best version of BB. Since Squeeze uses one $\lambda(\omega)$ map in the process, in theory $\lambda(\omega)$ would act as a lower bound for the performance of Squeeze, which in general is the case. Results with the TITAN RTX show a similar behavior than of the A100, once $n \ge 2^{13}$ and with $\lambda(\omega)$ as the lower bound. In the TITAN V, Squeeze at $\rho=16$ is faster than BB once $n \ge 2^{11}$. Surprisingly, for this GPU Squeeze produces an anomaly and manages to run slightly faster than the best version of lambda for $n \ge 2^{15}$ by a small margin. Possible causes of this anomaly are discussed in Section \ref{sec:conclusions}. From these plots, it becomes clear that the best absolute performance of Squeeze is at $\rho=16$.

\begin{figure*}[ht!]
\begin{center}
\includegraphics[scale=0.30]{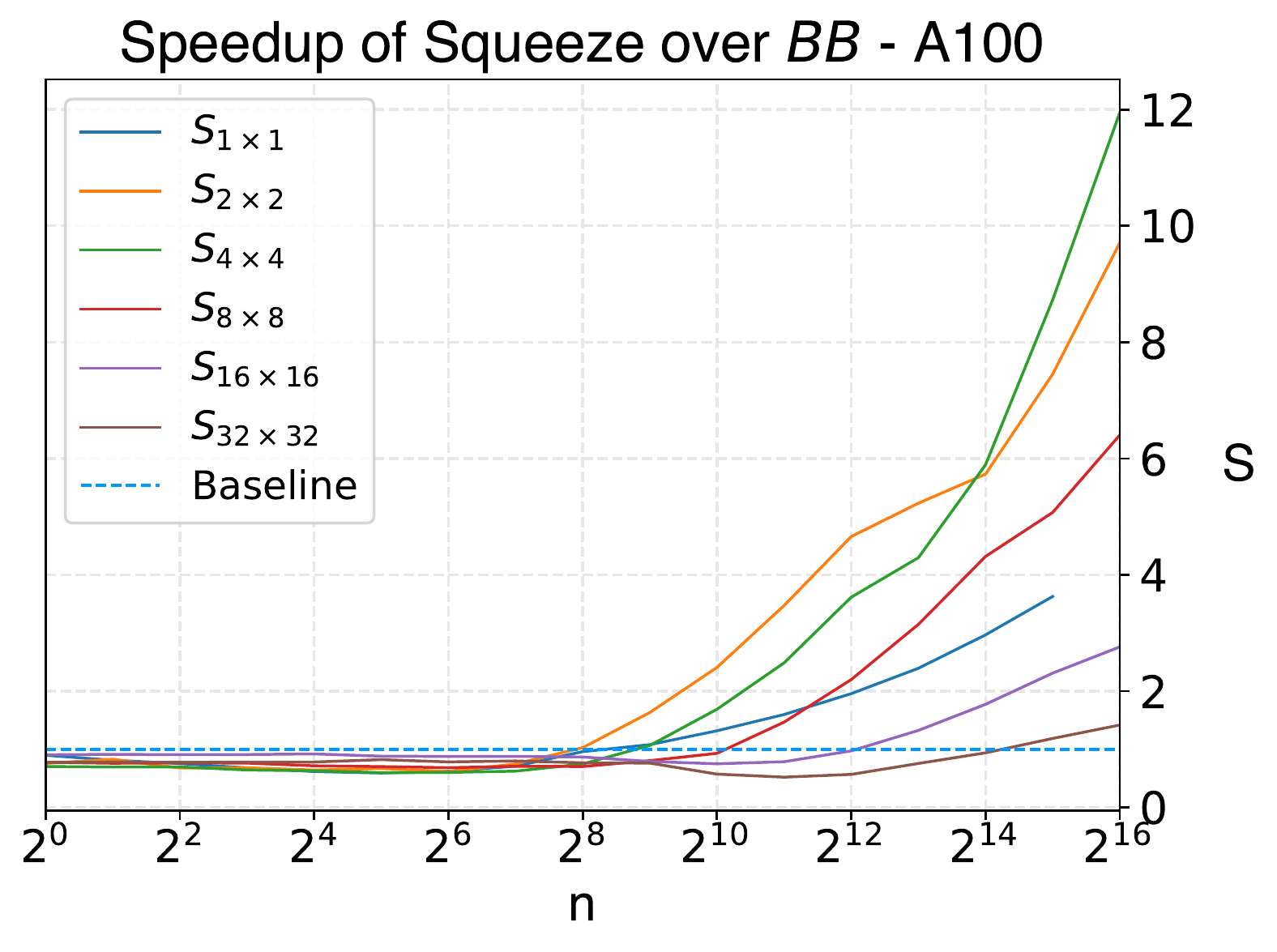}
\includegraphics[scale=0.30]{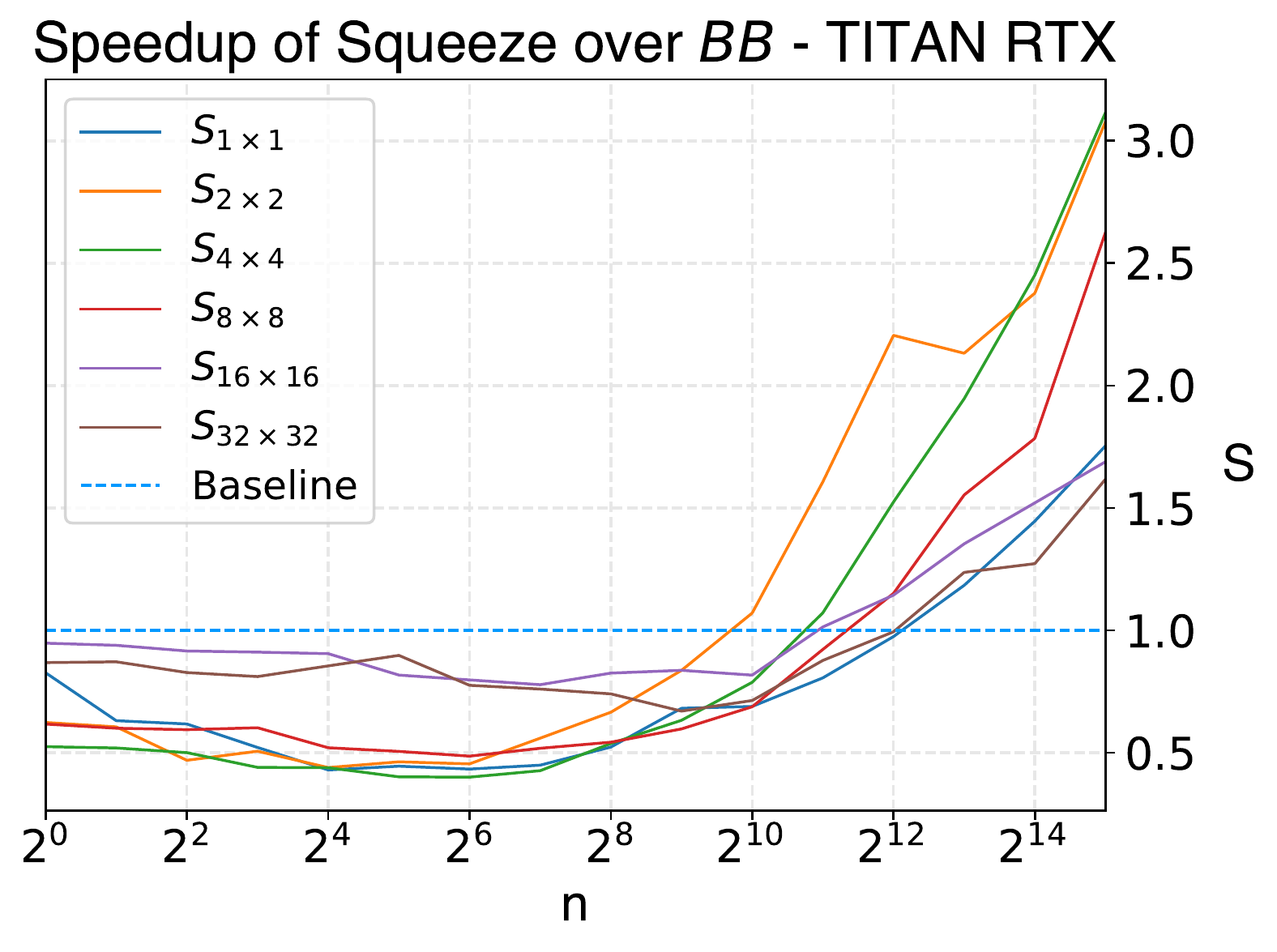}
\includegraphics[scale=0.30]{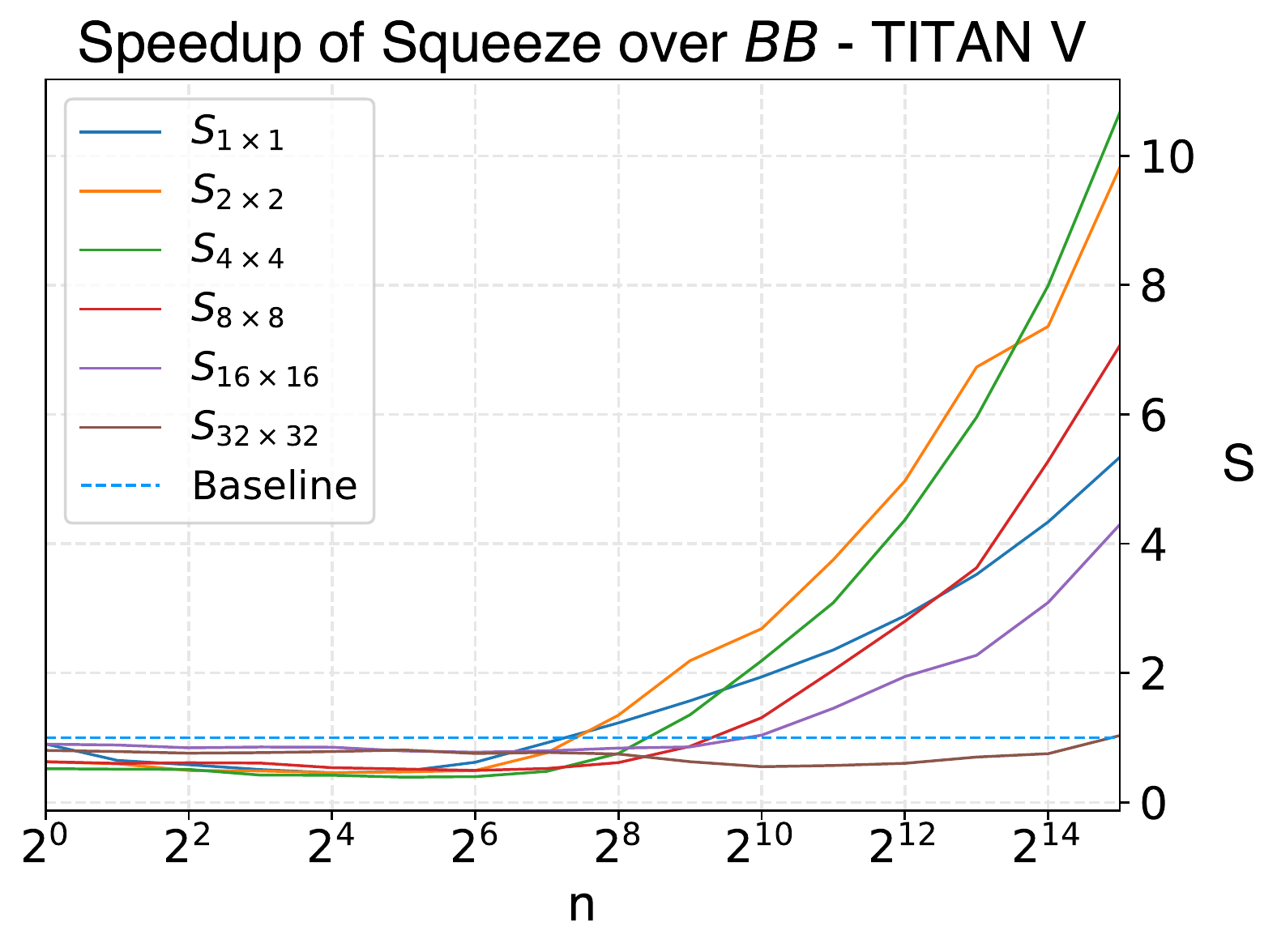}

\caption{Speedup of Squeeze over BB. Each curve is the speedup for a particular block size.
}
\label{fig:experimental-speedup}
\end{center}
\end{figure*}

Figure \ref{fig:experimental-speedup} presents the Speedup of Squeeze over BB at different block sizes $\rho$. 
Speedup curves of the NVIDIA A100 show that 
for $n \ge 2^8$ speedup starts to increase reaching a maximum of up to $\sim12\times$ with block sizes of $\rho = 1, 2, 4, 8$. For $\rho = 16, 32$ Squeeze reaches up to $\sim3.7\times$ of speedup. Thanks to the 40GB of GPU memory of the A100, it was possible to push the maximum problem size up to $2^{16}$, except for the curve of $S_{1x1}$ that could not reach the maximum size because of CUDA's grid size limits.
The TITAN RTX under-performed in comparison with the other GPUs for $n\leq2^{12}$. Past that value, speedup increases above $1$ for all values of $\rho$, reaching a top speedup of $\sim3.2\times$.
The TITAN V shows a similar pattern to the A100, with a top speedup of $\sim11\times$.

\begin{figure*}[ht!]
\begin{center}
\includegraphics[scale=0.30]{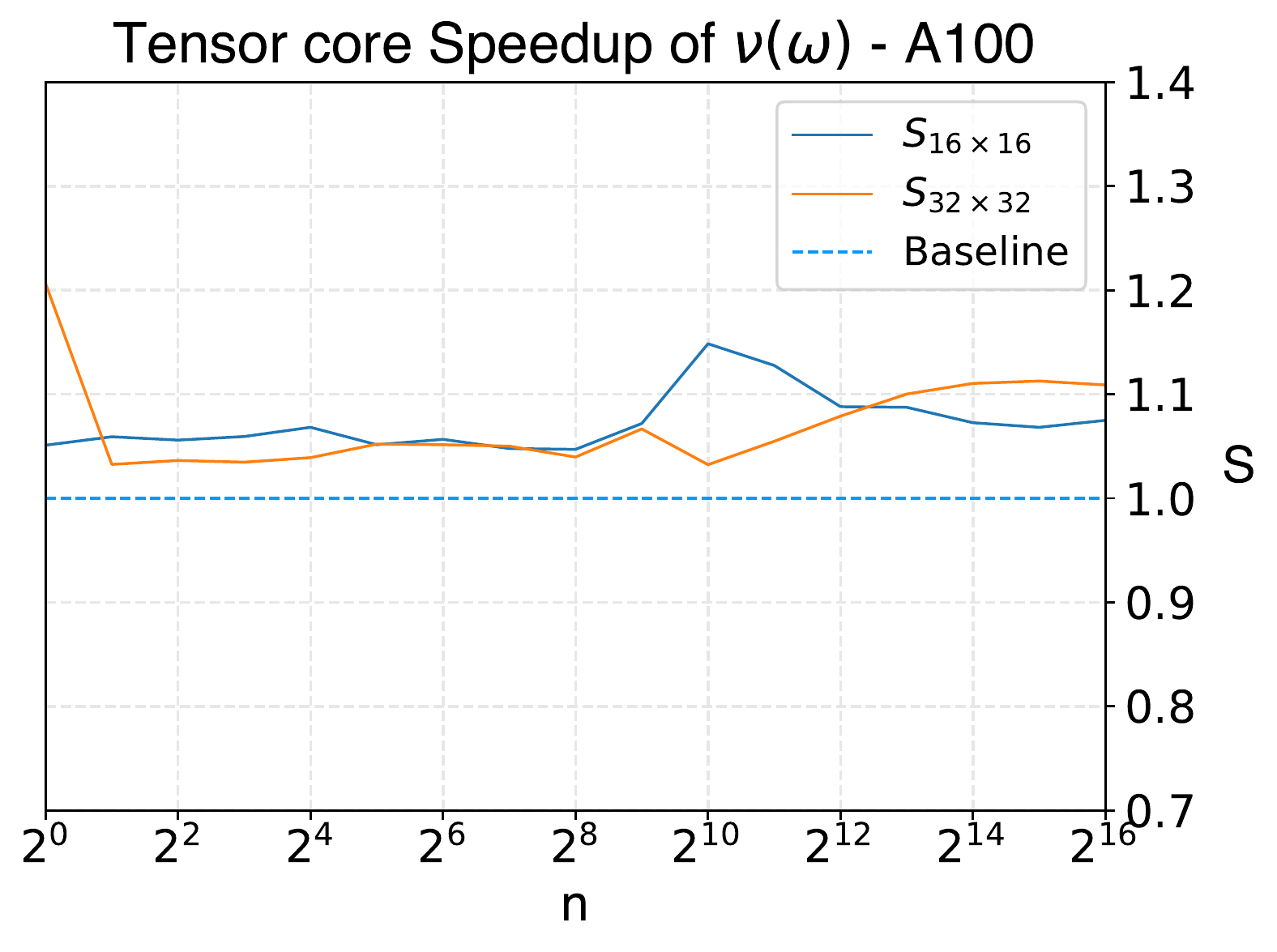}
\includegraphics[scale=0.30]{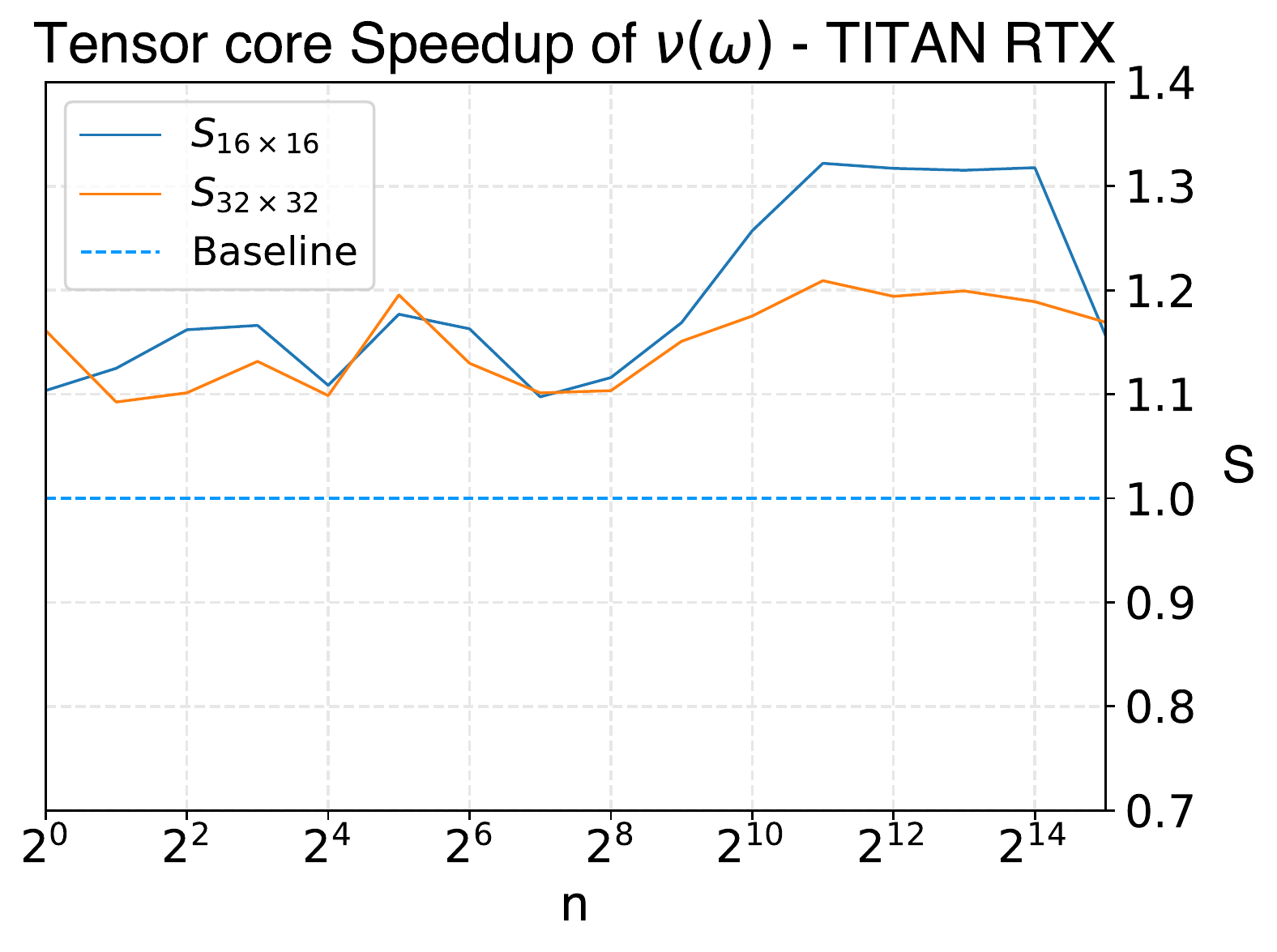}
\includegraphics[scale=0.30]{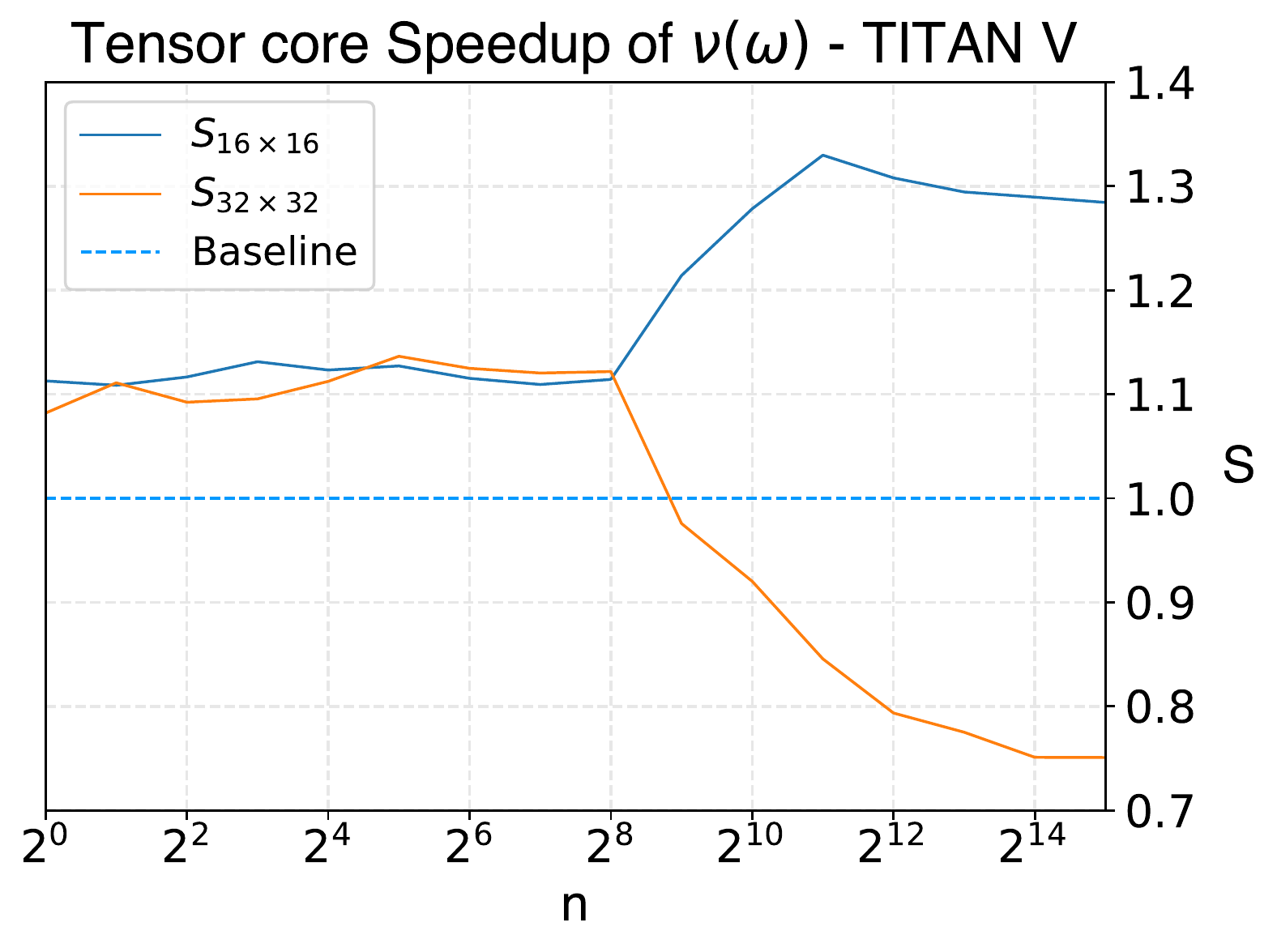}

\caption{The impact of using tensor cores vs not using them, for all three TCU generations (Volta, Turing and Ampere).}
\label{fig:tensor_perf}
\end{center}
\end{figure*}

Figure \ref{fig:tensor_perf} shows what is the 
performance contribution of the tensor cores. 
Results show a significant speedup across values of $n$, with the exception of $S_{32\times32}$ in TITAN V, where performance is negatively affected by using tensor cores, reaching $S~0.75\times$. The reason is unknown, but could be attributed to limitations in Volta architecture and the behavior of warps with first-gen tensor core units. The top speedup varies by GPU/Generation: on the A100 its $\sim1.11\times$, on the TITAN RTX is around $\sim1.2\times$ and on the TITAN V up to $\sim 1.3\times$. In any of the three cases, it is a significant extra-acceleration.

\subsection{Memory Reduction Factor}
Table \ref{table_memory} presents the total memory measured, as well as the memory reduction factor (MRF) by each approach to process a level $r=16$ Sierpinski Triangle across different block sizes. We recall that block sizes have a size-reduction effect on the fractal due to the block-level Squeeze approach. 
\begin{table}[ht!]
\caption{Total memory needed and memory reduction factor (MRF) for each approach on the Sierpinski triangle at $r=16$.}
\begin{center}
\begin{tabular}{| c | c | c | c |}
\hline
 $\rho$ & BB $|\ \lambda(\omega)$ & $\nu(\omega)$ & MRF\\
\hline
 $1\times1$ &  16GB & 0.16GB & $99.8\times$\\
 \hline
 $2\times2$	&  16GB & 0.21GB & $74.8\times$\\
\hline
 $4\times4$	&  16GB & 0.29GB & $56.1\times$\\
\hline
 $8\times8$	&  16GB & 0.38GB & $42.1\times$\\
\hline
 $16\times16$	&  16GB & 0.50GB & $31.6\times$\\
\hline
 $32\times32$	&  16GB & 0.68GB & $23.7\times$\\
\hline
\end{tabular}
\end{center}
\label{table_memory}
\end{table}

The table shows how small block sizes of $\rho = 1,2$ are close to a MRF of $100\times$, while larger block sizes gradually diminish the MRF factor because of the micro-fractals inside each block. Nonetheless, it is worth noticing that at $\rho = 16$ (Squeeze's best configuration), the MRF is $31.6\times$ which is a substantial improvement over BB and $\lambda(\omega)$.

Additional experiments showed that the 40GB of the A100 GPU allowed Squeeze to process fractals of up to level $r=20$, while BB and $\lambda(\omega)$ could only reach level $r=16$ before running out of memory. The $r=20$ achieved with Squeeze required from $\sim13$ to $\sim55$ GB of memory depending on the block size $\rho$. Processing a fractal of this size with a BB or $\lambda(\omega)$ approach would require $4096 GB$ of memory, this translates to a MRF of $\sim315\times$ for Squeeze.



\section{Discussion and Conclusions}
\label{sec:conclusions}
This work presented Squeeze: an approach for handling compact fractals efficiently on GPUs. By using an efficient compact representation, two benefits emerge: i) computation is employed only in the fractal elements and not in the entire embedding space, bringing speedup with respect to a bounding-box approach, and ii) it produces a significant reduction in memory usage, moving from expanded space to the Hausdorff dimension of the fractal. 
Tests with the A100 GPU provided a memory reduction factor (MRF) of up to $\sim99.8\times$, with a potential of $\sim315\times$ for level $r=20$. The MRF increases as the fractals becomes larger, meaning that future generation of GPUs could reach problem sizes that would be infeasible with traditional approaches that use the fractal's expanded embedded form.   

In terms of performance, tests with the Sierpinski triangle showed up to $12\times$ of speedup compared to a Bounding Box approach. Results also showed that the speedup, similar to the MRF, keeps increasing with the fractal size. When comparing Squeeze with the state of the art $\lambda(\omega)$ approach, we noted  $\lambda(\omega)$ is actually a performance lower bound. Although it is clear this lower-bound should exist as a general rule, we noted that one performance result with the Titan V GPU showed that Squeeze managed to run slightly faster than $\lambda(\omega)$. The cause of this anomaly might be related to internal properties of first-gen tensor cores or with the behavior of shared memory / memory accesses in Volta. Further research on this matter could give more insights.  

Adapting the computation of Squeeze to GPU tensor cores provided up to $30\%$ of extra performance compared to just using regular CUDA cores. Future improvements to tensor cores, both in quantity and performance, would provide an even higher  improvement. As future work, it would be useful to come up with a way to build arbitrary fractal structures by combining different NBB fractals at each scale level, as well as to extend Squeeze to support compact processing on 3D and higher-dimensional fractals. As a final conclusion, the proposed approach can be adopted by the community to accelerate simulations on NBB fractals and understand phenomena at larger scales.

\section*{Acknowledgement}
This research was supported by the Temporal research group (\url{http://temporal.uach.cl}), the ANID Fondecyt grant \#11180881 and the Patagón supercomputer from Universidad Austral de Chile (Fondequip EQM180042).
\bibliographystyle{elsarticle-num}
\typeout{}
\bibliography{manuscript}
\end{document}